\documentclass[onecolumn]{aa}
\usepackage[comma,authoryear]{natbib}
\usepackage[dvips]{graphicx}
\usepackage[english]{babel}
\usepackage[latin1]{inputenc}
\usepackage{latexsym}

\newcommand{\eqtoeq}[2]{Eqs.~(\ref{eq:#1})-(\ref{eq:#2})}

\newcommand{\eq}[1]{Eq.~(\ref{eq:#1})}
\newcommand{\eqs}[1]{Eqs.~(\ref{eq:#1})}
\newcommand{\Eq}[1]{Eq.~(\ref{eq:#1})}
\newcommand{\Eqs}[1]{Eqs.~(\ref{eq:#1})}
\newcommand{\fig}[1]{Fig.~\ref{fig:#1}}

\newcommand{\Tab}[1]{Table~\ref{tab:#1}}
\newcommand{\tab}[1]{Table~\ref{tab:#1}}

\newcommand{\app}[1]{Appendix~\ref{sect:#1}}

\newcommand{\sect}[1]{Sect.~\ref{sect:#1}}

\newcommand{\vsini}{v \cdot \sin i}
\newcommand{\kms}{\mathrm{km.s}^{-1}}
\newcommand{\Req}{R_\mathrm{eq}}
\newcommand{\Rp}{R_\mathrm{pol}}
\newcommand{\Lmax}{L_\mathrm{max}}
\newcommand{\Lres}{L_\mathrm{res}}
\newcommand{\Lmod}{L_\mathrm{mod}}
\newcommand{\Nr}{N_\mathrm{r}}

\newcommand{\vect}{\vec}

\newcommand{\grad}{\vect{\nabla}}

\newcommand{\lapl}{\Delta}

\renewcommand{\div}{\vect{\nabla} \cdot}
\renewcommand{\l}{\ell}
\renewcommand{\d}{\partial}
\renewcommand{\bar}{\underline}

\renewcommand{\S}{\mathcal{S}}
\newcommand{\G}{\mathcal{G}}

\renewcommand{\O}{\mathcal{O}}

\newcommand{\er}{\vect{e}_r}
\newcommand{\et}{\vect{e}_{\theta}}
\newcommand{\ep}{\vect{e}_{\phi}}

\newcommand{\ez}{\vect{e}_z}

\newcommand{\az}{\vect{a}_{\zeta}}
\newcommand{\at}{\vect{a}_{\theta}}
\newcommand{\ap}{\vect{a}_{\phi}}

\newcommand{\Hz}{H_{\zeta}}
\newcommand{\Ht}{H_{\theta}}

\newcommand{\Hzt}{H_{\zeta\theta}}

\newcommand{\sint}{\sin \theta}
\newcommand{\cost}{\cos \theta}
\newcommand{\cott}{\cot \theta}

\newcommand{\rz}{r_{\zeta}}
\newcommand{\rt}{r_{\theta}}
\newcommand{\rzz}{r_{\zeta\zeta}}
\newcommand{\rzt}{r_{\zeta\theta}}
\newcommand{\rtt}{r_{\theta\theta}}

\newcommand{\uz}{u^{\zeta}}
\newcommand{\ut}{u^{\theta}}
\newcommand{\up}{u^{\phi}}

\newcommand{\dz}{\partial_{\zeta}}
\newcommand{\dt}{\partial_{\theta}}
\newcommand{\dphi}{\partial_{\phi}}
\newcommand{\Dphi}{\mathrm{D}_{\phi}}
\newcommand{\dzz}{\partial^2_{\zeta\zeta}}
\newcommand{\dzt}{\partial^2_{\zeta\theta}}
\newcommand{\dtt}{\partial^2_{\theta\theta}}
\newcommand{\dpp}{\partial^2_{\phi\phi}}
\newcommand{\cz}{c_{\zeta}}

\newcommand{\ulm}{u_m^{\l}}
\newcommand{\ulmp}{u_m^{\l'}}
\newcommand{\vlm}{v_m^{\l}}
\newcommand{\vlmp}{v_m^{\l'}}
\newcommand{\wlm}{w_m^{\l}}
\newcommand{\wlmp}{w_m^{\l'}}

\newcommand{\Psilm}{\Psi_m^{\l}}
\newcommand{\Psilmp}{\Psi_m^{\l'}}
\newcommand{\blm}{b_m^{\l}}
\newcommand{\blmp}{b_m^{\l'}}
\newcommand{\Pilm}{\Pi_m^{\l}}
\newcommand{\Pilmp}{\Pi_m^{\l'}}

\newcommand{\Ylm}{Y^m_{\ell}}
\newcommand{\Ylmp}{Y^m_{\ell'}}

\newcommand{\Rlmp}{\vect{R}^m_{\ell'}}

\newcommand{\Slmp}{\vect{S}^m_{\ell'}}

\newcommand{\Tlmp}{\vect{T}^m_{\ell'}}
\newcommand{\Ntlm}{\dt Y_{\ell}^m}
\newcommand{\Ntlmp}{\dt Y_{\ell'}^m}
\newcommand{\Nplm}{\Dphi Y_{\ell}^m}
\newcommand{\Nplmp}{\Dphi Y_{\ell'}^m}

\newcommand{\Illm}[1]{\displaystyle I_{\ell\ell'}^m \left( #1 \right)}
\newcommand{\Jllm}[1]{\displaystyle J_{\ell\ell'}^m \left( #1 \right)}
\newcommand{\Jllmc}[1]{\displaystyle J\!c_{\ell\ell'}^m \left( #1 \right)}
\newcommand{\Kllm}[1]{\displaystyle K_{\ell\ell'}^m \left( #1 \right)}
\newcommand{\Kllmc}[1]{\displaystyle K\!c_{\ell\ell'}^m \left( #1 \right)}
\newcommand{\Lllm}[1]{\displaystyle L_{\ell\ell'}^m \left( #1 \right)}
\newcommand{\Mllm}[1]{\displaystyle M_{\ell\ell'}^m \left( #1 \right)}
\newcommand{\Mllmc}[1]{\displaystyle M\!c_{\ell\ell'}^m \left( #1 \right)}
\newcommand{\Nllm}[1]{\displaystyle N_{\ell\ell'}^m \left( #1 \right)}

\newcommand{\LNllm}[3]{\begin{array}{l} 
                       #1 L_{\ell\ell'}^m \\
                       #2 N_{\ell\ell'}^m
                       \end{array} \!\! \left( \displaystyle #3 \right)}

\newcommand{\MMcllm}[3]{\begin{array}{l} 
                       #1 M_{\ell\ell'}^m \\
                       #2 M\!c_{\ell\ell'}^m
                       \end{array} \!\! \left( \displaystyle #3 \right)}

\begin{document}

\title{Acoustic oscillations of rapidly rotating polytropic stars}

\subtitle{II.  Effects of the Coriolis and centrifugal accelerations}

\author{D. Reese \and F. Lignières  \and M. Rieutord}

\institute{
Laboratoire d'Astrophysique de Toulouse et Tarbes - UMR 5572 - 
Universit{\'e} Paul Sabatier Toulouse 3 - 
Observatoire Midi-Pyrénées, 14 avenue \'E. Belin,
31400 Toulouse, France}

\date{Received March 24, 2006; accepted Mai 12, 2006}

\offprints{D. Reese \\ \email{daniel.reese@ast.obs-mip.fr}}

\abstract
{With the launch of space missions devoted to asteroseismology (like COROT),
the scientific community will soon have accurate measurements of pulsation
frequencies in many rapidly rotating stars.}
{The present work focuses on the effects of rotation on pulsations of rapidly
rotating stars when both the Coriolis and centrifugal accelerations require a
non-perturbative treatment.}
{We develop a 2-dimensional spectral numerical approach which allows us to
compute acoustic modes in centrifugally distorted polytropes including the full
influence of the Coriolis force.  This method is validated through comparisons
with previous studies, and the results are shown to be highly accurate.}
{In the frequency range considered and with COROT's accuracy, we establish a
domain of validity for perturbative methods, thus showing the need for complete
calculations beyond $\vsini=50\,\kms$ for a $R = 2.3\,R_\odot$,
$M=1.9\,M_\odot$ polytropic star.  Furthermore, it is shown that the main
differences between complete and perturbative calculations come essentially
from the centrifugal distortion.} 
{}

\maketitle

\section{Introduction}

The study of rapidly rotating stars is a field in which there are many
unresolved questions.  The structure of these stars, the rotation profile, the
angular momentum transport and many other aspects are not well understood.  In
order to answer some of these questions, many different theoretical and
observational methods have been developed over the years.  For instance,
interferometry is starting to give clues as to the shape of these stars and
effects such as gravitational darkening \citep[\textit{e.g.}][]{Domiciano03,
Domiciano05,Peterson06}.  On the theoretical side, there exists a number of
numerical models which are progressively becoming more realistic
\citep[\textit{e.g.}][]{Roxburgh04,Jackson05,Rieutord05,Rieutord06}.  These
models can then be supplemented with asteroseismology which relates the
internal structure to observable stellar pulsations.  In order to fully exploit
these pulsations, it is necessary to accurately quantify how they are affected
by rotation. In the present work, we will show how this can be done for
acoustic pulsations in uniformly rotating polytropic stellar models.

Rotation has several effects on stars and their pulsations.  These result from
the apparition of two inertial forces, namely the centrifugal and the Coriolis
forces. The centrifugal force distorts the shape of the star and modifies its
equilibrium structure.  The Coriolis force intervenes directly in the
oscillatory motions.  Neither of these effects respect spherical geometry,
which means that the radial coordinate $r$ and the colatitude $\theta$ are no
longer separable.  As a result, pulsation modes cannot be described by a single
spherical harmonic as was the case for non-rotating stars.  In order to tackle
this problem, two basic approaches have been developed.  The first one is the
perturbative approach and applies to small rotation rates. In this approach,
both the equilibrium structure and the pulsation mode are the sum of a
spherical solution (or a single spherical harmonic), a perturbation, and a
remainder which is neglected.  The second approach consists in solving directly
the 2-dimensional eigenvalue problem fully including the effects of rotation.

Historically, the perturbative method has been applied to first, second and 
third order.  Previous studies include \citet{S81,GT90}, and \citet{DG92} for
second order methods and \citet{SGD98} and \citet{KCDPGD05} for third order
methods.  These have been applied to polytropic models \citep{S81} and then to
more realistic models.  There have also been some studies based on the
non-perturbative approach. Most non-perturbative calculations have focused on
the stability of neutron stars, r-modes and f-modes rather than on p-modes.
Some exceptions are \citet{C81,C84,C86,C89,C98,YE01} and \citet{EPHRC04}.

The present work aims at accurately taking into account the effects of rotation
on stellar acoustic pulsations, so as to be able to deduce asteroseismological
information from rapidly rotating stars.  Previous results are either
inaccurate or not valid for high enough rotation rates.  In order to achieve a
sufficient degree of precision, we used numerical methods which have already
proved to be highly accurate for other similar problems.  The present method is
a further development of the numerical method of \citet{Lignieres06} and
\citet{LRR06} (hereafter Paper~I) who used a spectral method \citep{Canuto88}
with a surface-fitting spheroidal coordinate system based on \citet{BGM98}. 
The spectral method itself has already been used for calculating inertial waves
in spherical shells \citep{Rieutord97} and gravito-inertial modes in a $1.5
\,M_\odot$ ZAMS star \citep{Dintrans00} both of which involve the
non-perturbative effects of the Coriolis force. Achieving high precision is of
great importance for interpreting present and future measurements of stellar
pulsations.  Furthermore, it provides a means to establish the domain of
validity of perturbative methods. Finally, this work can then be used as a
reference to validate future methods.  

The organisation of the paper is as follows: in the next section, the numerical
method is described in detail.  This section is followed by a series of
comparisons and tests which establish the accuracy of the results.  We then
proceed to discuss perturbative methods and their validity.  A conclusion and
outlooks follow.

\section{Formalism} 

The calculation of oscillation modes of rotating polytropes takes place in two
steps.  Firstly, an equilibrium model must be determined.  Secondly, this model
needs to be perturbed so as to give the eigenoscillations.

\subsection{Equilibrium model}

The equilibrium model is a self-gravitating uniformly rotating polytrope
described by the following equations in the rotating frame:
\begin{eqnarray}
\label{eq:polytrope}
P_o &=& K \rho_o^{\gamma}, \\
0 &=& -\grad P_o - \rho_o \grad \left( \Psi_o - \frac{1}{2}\Omega^2 s^2 \right),  \\
\lapl \Psi_o &=& 4\pi G\rho_o,
\end{eqnarray}
where $P_o$ is the pressure, $\rho_o$ the density, $K$ the polytropic constant,
$\gamma$ the polytropic exponent, $\Psi_o$ the gravitational potential, $s$ the
distance to the rotation axis and $G$ the gravitational constant.  One can also
introduce the polytropic index $N = 1/(\gamma-1)$ and a (pseudo-)enthalpy $h =
\int \mathrm{d}P/\rho = (1+N)P_o/\rho_o$. The pressure and density profiles are
then proportional to powers of this enthalpy: $P_o \propto h^{N+1}$, and
$\rho_o \propto h^N$.  A number of non-dimensional parameters also intervene
and characterise the polytropic model:
\begin{equation}
\label{eq:param}
\Lambda      = \frac{4\pi G \rho_c \Req^2}{h_c}, \qquad
\Omega_\star = \frac{\Omega \Req}{\sqrt{h_c}},   \qquad
\alpha       = \frac{\rho_c}{\left< \rho \right>}, \qquad
\varepsilon  = 1 - \frac{\Rp}{\Req},
\end{equation}
where quantities with the subscript ``c'' denote the equilibrium value at the
centre of the polytrope, $\Req$ and $\Rp$ are the equatorial and polar radii,
resp., and $\left< \rho \right> = 3M/4\pi \Req^3$ a pseudo-mean density. 
The method used to compute the equilibrium model is described in Paper~I.

\subsection{Perturbation equations}
We calculate adiabatic, inviscid oscillation modes using Eulerian perturbations
to the equilibrium quantities\footnote{The term ``perturbation'', which means a
small departure from equilibrium in this context, is not to be confused with
perturbation from the perturbative method, where it means a small departure
from the spherical case.}.  The linearised equations in the rotating frame
read:
\begin{eqnarray}
\d_t \rho            &=& - \div \left( \rho_o \vect{v} \right), \\
\rho_o \d_t \vect{v} &=& - \grad p + \rho \vect{g}_o - \rho_o \grad \Psi
                         - 2 \rho_o \vect{\Omega} \times \vect{v}, \\
\label{eq:adiabatic.energy}
\d_t p  - c_o^2 \d_t \rho &=& \frac{\rho_o N_o^2 c_o^2} {\|
                         \vect{g_o} \|^2} \vect{v} \cdot \vect{g}_o, \\
0                    &=& \lapl \Psi - 4 \pi G \rho,
\end{eqnarray}
where quantities with the subscript ``$o$'' denote equilibrium quantities and
those without any subscript Eulerian perturbations.  $\vect{g}_o$ is the
effective gravity, $c_o$ is the speed of sound, $\Gamma_1$ the adiabatic
exponent and $N_o$ the Brunt-V{\"a}is{\"a}l{\"a} frequency.  These are given by
the following formulas:
\begin{eqnarray}
\vect{g}_o &=& -\grad \left(\Psi_o - \frac{1}{2}\Omega^2 s^2 \right), \\
c_o^2      &=& \Gamma_1 P_o/\rho_o, \\
\Gamma_1   &=& \left( \frac{\d \ln p}{\d \ln \rho} \right)_{ad}, \\
N_o^2      &=& \vect{g_o} \cdot \left( -\frac{1}{\Gamma_1} \frac{\grad P_o}{P_o} +
\frac{\grad \rho_o}{\rho_o} \right).
\end{eqnarray}
It is worth noting that we have used the fluid's barotropicity in the definition
of $N_o$.

We can then put these equations in non-dimensional form using the following
transformations:
\begin{equation}
\begin{array}{*{19}{l}}
t          &=& T_r \bar{t},           & \qquad &
\rho       &=& \rho_c \bar{\rho},     & \qquad &
\vect{r}   &=& \Req \vect{\bar{r}},   & \qquad &
\vect{g}   &=& g_r \vect{\bar{g}},    & \qquad &
\vect{v}   &=& V_r \vect{\bar{v}},    \\
p          &=& P_r \bar{p},           & &
\Omega     &=& \omega_r \bar{\Omega}, & &
c_o        &=& V_r \bar{c}_o,         & &
N_o        &=& \omega_r \bar{N}_o,
\end{array}
\end{equation}
where:
\begin{equation}
\omega_r = T_r^{-1} = \left( 4 \pi G \rho_c \right)^{1/2}, \qquad
V_r      = \frac{\Req}{T_r}, \qquad
g_r      = \frac{\Req}{T_r^2} = 4\pi G \Req \rho_c, \qquad
P_r      = \frac{\rho_c \Req^2}{T_r^2}.
\end{equation}
It is important to note that $\Omega_\star$ and $\bar{\Omega}$ correspond to
two different dimensionless expressions of the rotation rate.  In order to go from
one expression to the other, one can use the following formula:
\begin{equation}
\Omega_\star = \bar{\Omega} \sqrt{\Lambda},
\end{equation}
where $\Lambda$ is given by \eq{param}.

If we assume a time dependence of the form $\exp(\lambda \bar{t})$, the
following generalised eigenvalue problem is obtained (we have dropped the
underlined notation):
\begin{eqnarray}
\lambda \rho            &=& -\vect{v} \cdot \grad \rho_o - \rho_o \div \vect{v}, \\
\lambda \rho_o \vect{v} &=& - \grad p + \rho \vect{g}_o  -\rho_o \grad \Psi
                            - 2 \rho_o \Omega \vect{e}_z \times \vect{v}, \\
\lambda p - \lambda c_o^2 \rho &=& \frac{\rho_o N_o^2 c_o^2}
                            {\| \vect{g_o} \|^2} \vect{v} \cdot \vect{g}_o, \\
0                       &=& \lapl \Psi - \rho.
\end{eqnarray}

\subsection{Change of variables}
In order to have solutions with a good numerical behaviour on the surface of the
star, we use the following variables:
\begin{equation}
\Pi =  \frac{p}{H^{N}}, \qquad
b   =  \frac{\rho}{H^{N-1}},
\end{equation}
where $H = h_o/h_c$ is a non-dimensional form of the enthalpy.  These choices
result from an analysis of the behaviour of the solution near the surface, based
on a ``generalised'' Frobenius study of the system of equations.  Although not
fully proved, this study gives the correct results in the spherical case (see
\app{Frobenius}).  It also leads to the following boundary condition on the
stellar surface:
\begin{equation}
\delta p/\rho_o = 0,
\label{eq:delta_p}
\end{equation}
where $\delta p$ is the Lagrangian pressure perturbation.  Not only is this
result in agreement with previous results, but it also specifies how fast $\delta p$
goes to zero near the stellar surface.  More details on this method are given in
\app{Frobenius}. This new choice of variables leads to the following set of
equations:
\begin{eqnarray}
\label{eq:continuity}
\lambda b          &=& - N \vect{v} \cdot \grad H - H \div \vect{v}, \\
\label{eq:Euler}
\lambda H \vect{v} &=& - H \left(\grad \Pi +\grad \Psi \right) 
                       + \grad H \left( -N \Pi + \frac{b}{\Lambda} \right)
                       - 2 \Omega H \ez \times \vect{v}, \\
\label{eq:energy}
\lambda \Pi         -  \lambda \frac{\Gamma_1}{(N+1)\Lambda} b &=&
                       \left( \frac{\Gamma_1}{\gamma} - 1 \right) 
                       \frac{\vect{v} \cdot \grad H}{\Lambda}, \\
\label{eq:Poisson}
0                  &=& \lapl \Psi - H^{N-1} b.
\end{eqnarray}

If $\Gamma_1 = \gamma$ then $N_o^2 = 0$ and the above system reduces
to:
\begin{eqnarray}
\label{eq:ad_continuity}
\lambda N \Lambda \Pi &=& - N \vect{v} \cdot \grad H - H \div \vect{v}, \\
\label{eq:ad_Euler}
\lambda  \vect{v}     &=&  - \grad \Pi -\grad \Psi - 2 \Omega  \ez \times
                          \vect{v}, \\
\label{eq:ad_Poisson}
0                     &=& \lapl \Psi - N \Lambda H^{N-1} \Pi.
\end{eqnarray}
This simplification occurs when the polytropic relation (\ref{eq:polytrope}) is
also the equation of state, a situation typical of white dwarfs or neutron
stars.  Furthermore, both $\Pi$ and $b$ become proportional to the Eulerian
perturbation of the enthalpy, thus justifying \textit{a posteriori} the choice
of these variables. As a result, apart from a few multiplicative factors, and
the lack of a dissipative force, this second set of equations corresponds to
those obtained by \citet{Y95}.

\subsection{Domains and boundary/interface conditions}

In order to complete the eigenvalue problem given by \eqtoeq{continuity}
{Poisson}, it is necessary to specify a number of boundary conditions. 
The basic requirements are that the solutions remain bounded at the surface
and at the centre of the star, and that the gravity potential goes to zero
at infinity.
 
At the centre of the star, the regularity conditions are classically expressed
in terms of spherical harmonics (see \eqs{regularity.even}
and~(\ref{eq:regularity.odd})).  By using the variables $\Pi$ and $b$ from the
generalised Frobenius study, the solution is naturally bounded on the star's
surface.  However, the use of these variables leads to a degeneracy between
\eqs{continuity}, (\ref{eq:energy}) and the radial component of \eq{Euler} on
the surface of the star.  This problem is remedied by replacing the radial
component of \eq{Euler} with its radial derivative on the surface.

It is also necessary to impose a boundary condition on the perturbation to the
gravity potential $\Psi$, in order to ensure that the potential goes to zero at
infinity.  Traditionally, this is done by doing a harmonic decomposition of
$\Psi$ and imposing the correct condition on each component.  However, such a
procedure becomes complicated on a spheroidal surface, and it is not certain
whether the decomposition of $\Psi$ will converge for highly flattened
configurations \citep{HES82}.  We therefore employ a different method based on
\citet{BGM98}.  It consists in adding a second domain $V_2$ which is bounded on
the inside by the star's surface and on the outside by a sphere of radius $r=2$
(which is twice the equatorial radius).  We solve Poisson's equation in this
domain and impose the correct boundary condition on its outer boundary (where
we can safely apply a harmonic decomposition).  On the inner boundary, it is
necessary to use interface conditions which ensure the continuity of $\Psi$ and
its radial derivative across the stellar surface.

\subsection{Spheroidal geometry}
\label{sect:spheroidal}

The next step in the calculations is the choice of a coordinate system based on
\citet{BGM98} for each domain.  In order to preserve spectral accuracy, the
system of coordinates in the first domain needs to fit the surface of the star,
and provide a non-singular transformation in the centre.  As in Paper~I and
\citet{Rieutord05}, we choose the following definition for the radial
coordinate $\zeta$, which ensures a good convergence of the numerical method:
\begin{equation}
r(\zeta,\theta) = (1-\varepsilon)\zeta+\frac{5\zeta^3-3\zeta^5}{2}
                 \left( R_s(\theta) - 1 + \varepsilon \right),
\end{equation}
where $\varepsilon$ is the flatness given by \eq{param}, $(r(\zeta,\theta),
\theta,\phi)$ are the spherical coordinates corresponding to the point
$(\zeta,\theta,\phi)$, and $R_s(\theta)$ is the surface of the star.  By
setting $\zeta=1$, one obtains $r(1,\theta) = R_s(\theta)$, and the centre
$r=0$ is given by $\zeta = 0$.

In second domain, we used the following definition:
\begin{equation}
r(\zeta,\theta) = 2\varepsilon + (1-\varepsilon) \zeta 
                 + \left( 2\zeta^3 - 9\zeta^2+12\zeta-4\right)
                   \left( R_s(\theta) - 1 - \varepsilon \right),
\end{equation}
where $\zeta \in [1,2]$.  This mapping is chosen so as to insure the continuity
of $r$ and $\rz$ across the boundary $\zeta = 1$, and so that the surface given
by $\zeta=2$ corresponds to the sphere $r=2$ ($\rz$ denotes $\dz r$).

\begin{figure}[ht!]
\begin{tabular}{ll}
\parbox[][12cm][t]{12cm}
{\includegraphics[width=12cm,height=12cm]{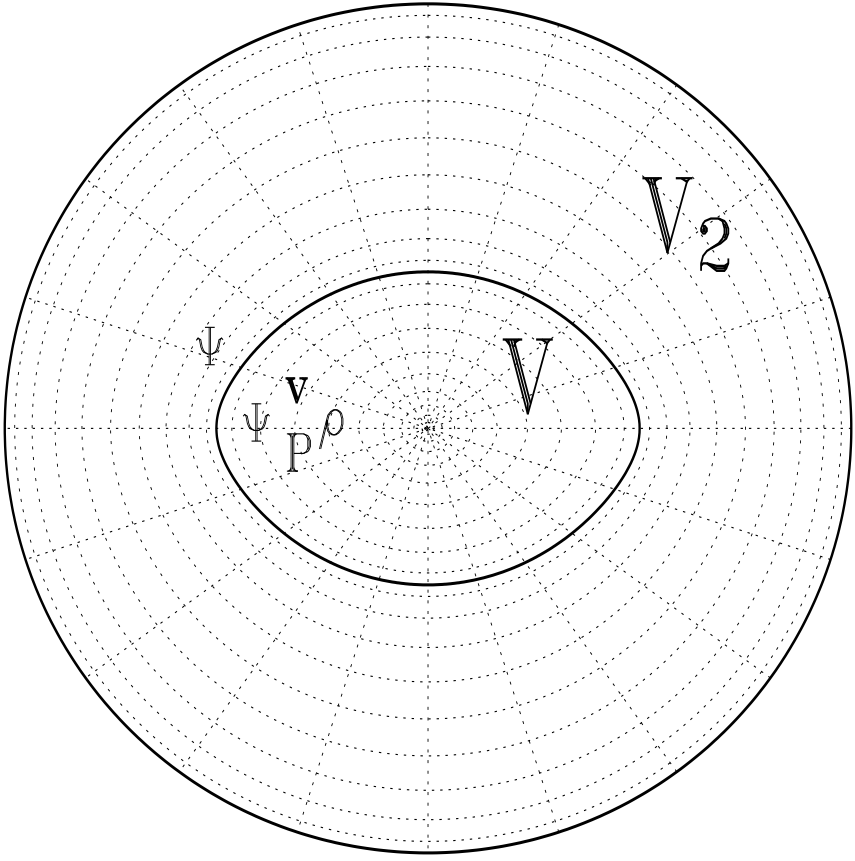}} &
\parbox[][12cm][t]{5.5cm}
{\caption{Coordinate system used in computing the equilibrium model of the star
and the pulsation modes.  The domain $V$ corresponds to the star itself (which
in this case is a $N = 3$ polytrope at $84 \%$ of the breakup rotation rate). 
The domain $V_2$ encompasses the star, its outer limit being a sphere of radius
$r=2$ (twice the equatorial radius).  The dotted lines correspond to $\zeta =
0.1,\,0.2\, ... 0.9$ in the domain $V$, $\zeta = 1.1,\,1.2\, ... 1.9$ in the
domain $V_2$ and  $\theta = 0^{\circ},\, 18^{\circ},\,... 342^{\circ}$.  The
centre of the star corresponds to $\zeta=0$, the star's surface (the boundary
between $V$ and $V_2$) to $\zeta=1$, and the outer boundary to $\zeta=2$. The
continuity equation, Euler's equation, the energy equation and Poisson's
equation are solved in the first domain.  The letters $\Psi$, $\vect{v}$, $P$
and $\rho$ in the domain $V$ show that these variables intervene in the first
domain.  In the second domain, only Poisson's equation is solved, and only the
perturbation of the gravity potential $\Psi$ intervenes.  This is represented
by the letter $\Psi$ in the domain $V_2$.}
\label{fig:domains}}
\end{tabular}
\end{figure}

Once the coordinate system has been established, it is also necessary to choose
a set of vectors as a basis.  We define the following vectors, which are
derived from the natural covariant basis $(\vect{E}_{\zeta}, \vect{E}_{\theta},
\vect{E}_{\phi})$ (defined as $\vect{E}_i=\d_i\vect{r}$):
\begin{equation}
\begin{array}{lllll}
\vect{a}_{\zeta}  &=& \displaystyle \frac{\zeta^2}{r^2 \rz} \vect{E}_{\zeta} 
                  &=& \displaystyle \frac{\zeta^2}{r^2} \er, \\
\vect{a}_{\theta} &=& \displaystyle \frac{\zeta}{r^2 \rz} \vect{E}_{\theta} 
                  &=& \displaystyle \frac{\zeta}{r^2 \rz} \left( \rt \er + r \et
                      \right), \\
\vect{a}_{\phi}   &=& \displaystyle \frac{\zeta}{r^2 \rz \sint} \vect{E}_{\phi}
                  &=& \displaystyle \frac{\zeta}{r \rz} \ep,
\end{array}
\end{equation}
where $(\er,\et,\ep)$ are the usual spherical vectors.  The vectors
$(\az,\at,\ap)$  have been chosen so that they become $(\er,\et,\ep)$ in the
spherical limit.  Using this base of vectors, we can then express the velocity
field as follows:
\begin{equation}
\vect{v} = \uz \az + \ut \at + \up \ap.
\end{equation}
With these definitions, it is now possible to give an explicit expression of the
oscillation equations:


\begin{eqnarray}
\label{eq:spheroidal.continuity}
\lambda b &=& 
 -\frac{N\zeta^2}{r^2 \rz}
  \left[  \Hz \uz + \frac{\Ht \ut}{\zeta}  \right]
              - \frac{\zeta^2 H}{r^2 \rz}\left[
                \dz \uz 
              + \frac{2\uz}{\zeta} 
              + \frac{\dt \ut}{\zeta}           
              + \frac{\cot \theta \ut}{\zeta} 
              + \frac{\dphi \up}{\zeta \sin \theta}  
               \right], \\
\noalign{\smallskip} 
\label{eq:spheroidal.Euler1}
\lambda  \left[ \frac{\zeta^2 H \rz \uz}{r^2} 
        +\frac{\zeta H \rt \ut }{r^2}\right] &=&
         \frac{2 \Omega H \zeta \sint \up}{r}  
        - H \left( \dz \Pi + \dz \Psi \right)
        +\Hz \left( \frac{b}{\Lambda} - N \Pi \right), \\
\noalign{\smallskip} 
\label{eq:spheroidal.Euler2}
\lambda   \left[ \frac{\zeta^2 \rt \uz}{r^2} + 
          \frac{\zeta(r^2+\rt^2) \ut}{r^2\rz} \right] &=&
          \frac{2 \Omega \zeta (\rt \sint + r \cost)\up}{r\rz} 
         -\dt \Pi
         -\dt \Psi 
         +\frac{\Ht}{H} \left( \frac{b}{\Lambda} - N \Pi \right), \\
\noalign{\smallskip} 
\label{eq:spheroidal.Euler3}
\lambda   \frac{\zeta\up}{\rz} &=&
           -\frac{2 \Omega \zeta^2 \sint\uz}{r}
           -\frac{2 \Omega \zeta(\rt \sint + r \cost)\ut}{r\rz}
           -\frac{\dphi \Pi}{\sint}
           -\frac{\dphi \Psi}{\sint}, \\
\noalign{\smallskip} 
\label{eq:spheroidal.energy}
\lambda \left( \Pi - \frac{\Gamma_1 b}{(N+1)\Lambda} \right) &=& 
\frac{\zeta^2}{\Lambda r^2 \rz}
\left( \frac{\Gamma_1}{\gamma} - 1 \right) 
\left[ \Hz\uz + \frac{\Ht\ut}{\zeta} \right], \\
\noalign{\smallskip} 
\label{eq:spheroidal.Poisson}
0 &=& \frac{r^2 + \rt^2}{r^2 \rz^2}  \dzz \Psi
+\cz  \dz \Psi 
-\frac{2\rt}{r^2 \rz} \dzt \Psi
+\frac{1}{r^2} \lapl_{\theta \phi} \Psi
- H^{N-1} b,
\end{eqnarray}
where:
\begin{eqnarray}
\cz &=& \frac{1}{r^2 \rz^3} \left( 2 \rz \rt \rzt - r^2 \rzz - \rz^2 \rtt 
+ 2 r\rz^2 - \rt^2 \rzz -  \rz^2 \rt \cott \right), \\
\lapl_{\theta\phi} &=& \dtt + \cott \dt + \frac{1}{\sin^2 \theta}\dpp.
\end{eqnarray}

\Eq{spheroidal.continuity} is the continuity equation,
\eqtoeq{spheroidal.Euler1}{spheroidal.Euler3} are Euler's equations,
\eq{spheroidal.energy} corresponds to the adiabatic energy equation and
\eq{spheroidal.Poisson} is Poisson's equation.  Euler's equations have been
used in their covariant rather than contravariant form, as it is advantageous
from a numerical point of view.  In order to understand this, it is helpful to
bear in mind that $\Ht/H$ converges towards $\Hzt/\Hz$ on the stellar surface,
whereas $\Hz/H$ is unbounded (since $\Hz \neq 0$ on the surface).  As a result,
the radial component of Euler's equation reduces to $b = \Lambda N \Pi$ on the
surface (which incidentally is already implied by a linear combination of the
energy and continuity equations), whereas the two other components retain
useful information on the surface.  If the equations where in their
contravariant form, than the $\theta$ component of Euler's equation would also
reduce to $b = \Lambda N \Pi$ on the stellar surface since $\Hz$ would appear
in this equation, thus preventing the possibility of dividing by $H$.  It would
then be necessary to also replace this equation by a supplementary boundary
condition which would provide the information already contained in the
covariant form.  This system of equations applies in the first domain, except
for Poisson's equation which is used in both domains (of course, the density
perturbation no longer appears in the second domain).

\subsection{Numerical Method}

In order to solve \eqtoeq{spheroidal.continuity}{spheroidal.Poisson}, we
project these equations onto the spherical harmonics \citep{Rieutord87}.  This
is done in two steps (\textit{c.f.} Paper~I). First of all, the different
unknowns are expressed in terms of a sum over the spherical harmonics. 
Explicitly, we obtain:
\begin{eqnarray}
\label{eq:harmonic.density}
b    &=& \sum_{\l'=|m|}^{\infty} \blmp \Ylmp, \\
\Pi  &=& \sum_{\l'=|m|}^{\infty} \Pilmp \Ylmp, \\
\Psi &=& \sum_{\l'=|m|}^{\infty} \Psilmp \Ylmp, \\
\label{eq:harmonic.velocity}
\vect{v} &=& \sum_{\l'=|m|}^{\infty} \ulmp \Rlmp + \vlmp \Slmp + \wlmp \Tlmp,
\end{eqnarray}
where $\Ylmp$ is the spherical harmonic of degree $\l'$ and azimuthal order $m$
and $\blmp$, $\Pilmp$ etc. are radial functions that need to be determined, and
which only depend on $\zeta$. The equilibrium model is axisymmetric meaning
that the variable $\phi$ is not coupled to the two others variables. 
Therefore, there is no summation over the azimuthal order $m$ in these
expressions.  However, $\zeta$ and $\theta$ are not separable since the star
does not respect spherical symmetry.  As a result, it is necessary to sum over
the harmonic degree $\l'$.

$\Rlmp$, $\Slmp$, and $\Tlmp$ are defined as follows:
\begin{eqnarray}
\Rlmp &=& \Ylmp \az, \\
\Slmp &=& \Ntlmp \at + \Nplmp \ap,\\
\Tlmp &=& \Nplmp \at - \Ntlmp \ap,\\
\Dphi & \equiv & \frac{\dphi}{\sint}.
\end{eqnarray}
It is worth noting that $\Rlmp$, $\Slmp$, and $\Tlmp$ are \emph{not} the usual
vectorial spherical harmonics because $(\az,\at,\ap)$ is not the same as
$(\er,\et,\ep)$. However, in the spherical limit, they will become the usual
spherical harmonics.  An explicit expression for each component of the velocity
reads:
\begin{eqnarray}
\uz &=& \sum_{\l'=|m|}^{\infty} \Ylmp  \ulmp, \\
\ut &=& \sum_{\l'=|m|}^{\infty} \Ntlmp \vlmp + \Nplmp \wlmp, \\
\up &=& \sum_{\l'=|m|}^{\infty} \Nplmp \vlmp - \Ntlmp \wlmp.
\end{eqnarray}

Once the unknown quantities have been expressed this way, the next step is to
project the equations themselves onto the spherical harmonic basis.  
\Eqs{spheroidal.continuity}, (\ref{eq:spheroidal.Euler1}),
(\ref{eq:spheroidal.energy}) and (\ref{eq:spheroidal.Poisson}) are multiplied
by $\left\{ \Ylm \right\}^*$ and integrated over $4\pi$ steradians.  For each
harmonic degree $\l$ of $\left\{ \Ylm \right\}^*$, a different equation is
obtained. The remaining equations are obtained from \eqs{spheroidal.Euler2}
and~(\ref{eq:spheroidal.Euler3}) in a more complicated manner.  We compute the
integral over $4\pi$ radians of $\left\{ \mbox{\eq{spheroidal.Euler2}} \right\}
\left\{ \Ntlm \right\}^* + \left\{ \mbox{\eq{spheroidal.Euler3}} \right\}
\left\{ \Nplm \right\}^*$ and $\left\{ \mbox{\eq{spheroidal.Euler2}} \right\}
\left\{ \Nplm \right\}^* - \left\{ \mbox{\eq{spheroidal.Euler3}} \right\}
\left\{ \Ntlm \right\}^*$.  This operation corresponds to what would be done in
the spherical case (\textit{i.e.} a projection onto the vectorial spherical
harmonics).  As a result, the system thus obtained reduces to the classical
uncoupled system of equations in the spherical limit.  In general, however,
this set of equations is a highly coupled system of ordinary differential
equations in terms of the radial coordinate $\zeta$, the solution of which
gives the unknown radial functions (see \app{harmonic}).

In order to solve this system numerically, we first begin by using a finite
number of spherical harmonics $\Lmax$.  The equations are then discretised onto
a Gauss-Lobatto collocation grid of $\Nr+1$ points, based on the Chebyshev
polynomials.  This results in an algebraic system of the form $\mathcal{A}v =
\lambda \mathcal{B} v$ in which $\mathcal{A}$ and $\mathcal{B}$ are numerically
determined square matrices.  The eigensolutions $(\lambda,v)$ of this
system correspond to the frequencies and pulsation modes of the star.  They are
determined iteratively through the Arnoldi-Chebyshev algorithm
\citep[\textit{e.g.}][]{Chatelin88}. The coefficients of matrices
$\mathcal{A}$ and $\mathcal{B}$ are computed using an
equilibrium model with a harmonic resolution $\Lmod$ and a Chebyshev (radial)
resolution of $\Nr+1$. They are calculated using the coupling
integrals given in \app{harmonic}.  This is achieved through Gauss' quadrature
method with $\Lres$ points.  Typical values for the different resolutions are:
$\Nr=60$, $\Lmod = 50$, $\Lmax = 80$, $\Lres = 230$.

At this point, we can write the boundary condition on the gravitational
potential and the regularity conditions at the centre.  The boundary condition
is applied along the surface $r_{ext}=2$ (or $\zeta=2$) on each harmonic
component of the gravitational potential perturbation \citep{HRW66}:
\begin{equation}
\frac{1}{1-\varepsilon}\frac{\mathrm{d}\Psilm}{\mathrm{d}\zeta}+\frac{\l+1}{r_{ext}}\Psilm=0.
\end{equation}
The regularity condition depends on the parity of $\l$ in a solution.  Thanks
to star's equatorial symmetry, modes will either be described by a sum of
spherical harmonics with even degrees or odd degrees\footnote{The toroidal
components $\wlm$ have the opposite parity with respect
to the other components.} (see \sect{symmetries}).  For modes with even
harmonics, we apply the following condition at $r=0$ (or $\zeta=0$):
\begin{equation}
\label{eq:regularity.even}
\frac{\mathrm{d}\Psilm}{\mathrm{d}\zeta} = 0, \qquad
\frac{\mathrm{d}\Pilm}{\mathrm{d}\zeta} = 0, \qquad
\frac{\mathrm{d}\blm}{\mathrm{d}\zeta} = 0, \qquad
\ulm = 0, \qquad
\vlm = 0, \qquad
\wlm = 0.
\end{equation}
The other modes follow the condition:
\begin{equation}
\label{eq:regularity.odd}
\Psilm = 0, \qquad
\Pilm = 0, \qquad
\blm = 0, \qquad
\frac{\mathrm{d}\ulm}{\mathrm{d}\zeta} = 0, \qquad
\frac{\mathrm{d}\vlm}{\mathrm{d}\zeta} = 0, \qquad
\frac{\mathrm{d}\wlm}{\mathrm{d}\zeta} = 0.
\end{equation}

\subsection{Mode classification and symmetries}
\label{sect:symmetries}

A number of useful pieces of information can be deduced from the various
symmetries present in the system.  These help with mode classification, reduce
numerical demand and explain certain properties which were observed in
perturbative calculations.

The first and most obvious symmetry stems from the fact that the equilibrium
model is axisymmetric.  This implies that modes will have a well defined
azimuthal order $m$ (as explained earlier on). A second equally obvious
symmetry results from the star being symmetric with respect to the equatorial
plane.  This leads to oscillation modes which are either symmetric or
antisymmetric with respect to the equatorial plane and which are called even or
odd, respectively.  In terms of spherical harmonics, even modes are made up of
harmonic components such that $\l+m$ is even, except for the toroidal component
of the velocity field in which $\l+m$ is odd.  Odd modes correspond to the
opposite situation.  From a numerical point of view, eigensolutions are
described with half as many components as solutions with no particular parity.

There are two more symmetries which are a little more subtle than the previous
ones. The first one only applies if the Coriolis force is neglected and only
shows up in the rotating frame.  In this situation, only even powers of the
rotation rate show up in the equilibrium and pulsation equations.  As a result,
a given mode will also only depend on $\Omega^2$ and will be a solution for the
rotation rates $\Omega$ and $-\Omega$.  When this symmetry is combined with the
next one, then for a given multiplet, modes with azimuthal orders $m$ and $-m$
have the same frequency, as was already pointed out in Paper~I.

The last symmetry applies even with the Coriolis force and for both rotating
and non-rotating frames.  Let us consider a solution $(\omega, \rho, P,
\vect{v}, \Psi, \Omega, m)$ (we include the rotation rate and the azimuthal
order for the sake of clarity) and denote by $\S$ the operator which gives the
mirror image with respect to the meridian passing through $\phi =
0$\footnote{In spherical coordinates, $\S$ is defined as follows for a scalar
quantity: $\S A(r,\theta,\phi) = A(r,\theta,-\phi)$.  For a vector field it
takes on the definition: $\S \vect{V}(r,\theta,\phi) = V_r(r,\theta,-\phi) \er
+ V_{\theta} (r,\theta,-\phi) \et - V_{\phi}(r,\theta,-\phi) \ep$.}. We then
find that $(\omega, \S\rho, {\S}P, \S\vect{v}, \S\Psi, -\Omega, -m)$ is also a
solution (this is not to be confused with the previous symmetry for which
$(\omega, \rho, P, \vect{v}, \Psi, -\Omega, m)$ was the corresponding
solution). This symmetry was pointed out by \citet{C89}, however some of its
consequences on perturbative calculations were not fully appreciated at the
time.  Let us consider a perturbative description of two frequencies with the
same radial order and harmonic degree but with opposite azimuthal orders.  We
will obtain expressions of the following form:
\begin{eqnarray}
\label{eq:coeff.sym}
\omega_{n,\,{\l},\,m}(\Omega)  &=& \omega_{n,\,{\l},\,m}^0 + \omega_{n,\,{\l},\,m}^1 \Omega 
                                 + \omega_{n,\,{\l},\,m}^2 \Omega^2 + ... + \O\left(\Omega^k\right), \\
\omega_{n,\,{\l},\,-m}(\Omega) &=& \omega_{n,\,{\l},\,-m}^0 + \omega_{n,\,{\l},\,-m}^1 \Omega 
                                 + \omega_{n,\,{\l},\,-m}^2 \Omega^2 + ... + \O\left(\Omega^k\right),
\end{eqnarray}
where $\omega_{n,\,{\l},\,m}^j$ is the $j^\mathrm{th}$ perturbative coefficient
of $\omega_{n,\,{\l},\,m}(\Omega)$. If we apply the symmetry, we find that
$\omega_{n,\,{\l},\,m}(\Omega) = \omega_{n,\,{\l},\,-m} (-\Omega)$.  By
equating like powers of  $\omega_{n,\,{\l},\,-m} (-\Omega)$ with
$\omega_{n,\,{\l},\,m} (\Omega)$, we find that $\omega_{n,\,{\l},\,m}^j =
(-1)^j \omega_{n,\,{\l},\,-m}^j$.  Therefore $\omega_{n,\,{\l},\,m}^j$ will be
an even function of $m$ when $j$ is even and an odd function of $m$ when $j$ is
odd.  This explains why the second order coefficients of \citet{DG92} were
polynomials in $m^2$, and is also found at third order (Goupil, private
communication).  This symmetry can also be used to increase the accuracy of
least squares estimates of the coefficients based on non-perturbative
calculations (see \sect{least.squares}).

\section{Analysis of the accuracy of the results}

In order to check whether the results presented here are correct, it is
important to do a number of internal tests and comparisons with previous
studies.   We first begin by discussing the accuracy of the underlying
polytropic models.  This is then followed by a series of comparisons with other
studies.  In the two first comparisons, the previous results have a limited
accuracy, therefore only allowing a qualitative evaluation. The next two
comparisons are with very accurate results, thus allowing a quantitative
evaluation of the precision of the present results.  These are then followed by
a test based on the variational principle and an analysis of the sensitivity of
the results to the parameters used in the numerical method. Finally, we
conclude by estimating the overall accuracy of the results.

\subsection{Accuracy of the polytropic models}
\label{sect:model.precision}

There are several different tests which give an idea of the accuracy of the
polytropic models.  One way is by looking at the effects of different input
parameters, such as the radial or harmonic resolution, on various
non-dimensional parameters like those in \eq{param}.  For non-rotating models,
these non-dimensional parameters can be compared with those given in
\citet{Seidov04}. In \tab{Seidov}, we give such a comparison, which shows that
it is possible to correctly obtain $6$ digits after the decimal point.
\Tab{param.rota} contains $\alpha$ and $\Lambda$ for an $N=3$ polytrope
rotating at $0.59\,\,\Omega_K$.  This table shows the strong influence of $\Nr$
and the need for a sufficient radial resolution.  It also suggests a precision
of $6$ digits after the decimal point, if we compare the values for $\Nr=50$ and
$\Nr=60$. 

\begin{table}[ht]
\caption{Non-dimensional parameters of a non-rotating $N=3$ polytrope. 
$L_\mathrm{max} = 50$ for all the calculations. It is difficult to accurately
obtain the $7^\mathrm{th}$ digit after the decimal point, in comparison with
the values of \citet{Seidov04}.}
\label{tab:Seidov}
\begin{tabular}{*{3}l}
\hline
\hline
$\Nr$ & $\alpha$ & $\Lambda$ \\
\hline
50 &
54.182 480 87 & 
47.566 520 74 \\ 
60 &
54.182 481 06 & 
47.566 520 85 \\
100 &
54.182 480 87 & 
47.566 520 74 \\
\multicolumn{1}{l}{\citet{Seidov04}} & 
54.182 481 11 & 
47.566 520 88 \\
\hline
\end{tabular}
\end{table}

\begin{table}[ht]
\caption{Same as \tab{Seidov} but for an $N=3$ polytrope rotating
at $0.59\,\,\Omega_K$.  The radial resolution $\Nr$ has a stronger
effect on the values of $\alpha$ and $\Lambda$ than the harmonic resolution
$\Lmax$.}
\label{tab:param.rota}
\begin{tabular}{*{4}l}
\hline
\hline
$\Nr$ & $\Lmax$ & $\alpha$ & $\Lambda$ \\
\hline
50 & 16 & 
81.108 265 69 &
63.025 583 86 \\
20 & 50 & 
81.108 444 82 &
63.025 591 55 \\
50 & 50 & 
81.108 249 13 &
63.025 575 39 \\
50 & 60 & 
81.108 249 08 &
63.025 575 36 \\
60 & 50 & 
81.108 249 38 &
63.025 575 52 \\
60 & 60 & 
81.108 249 38 &
63.025 575 52 \\
\hline
\end{tabular}
\end{table}

\begin{figure}[ht]
\begin{tabular}{ll}
\parbox[][7cm][t]{10cm}
{\includegraphics[width=10cm]{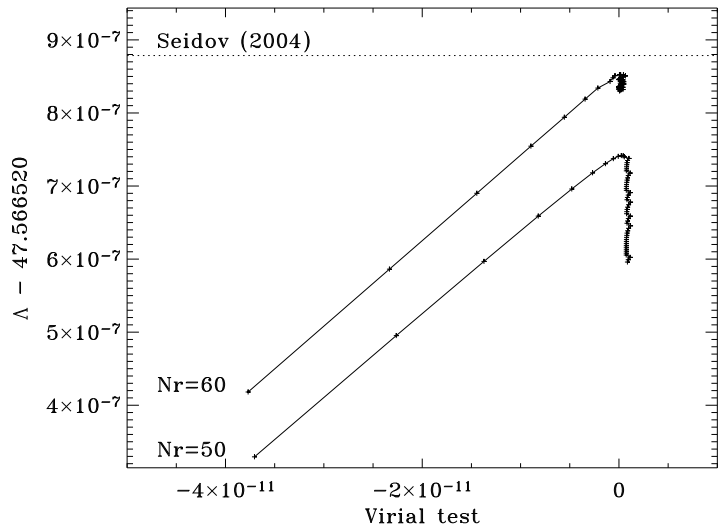}} &
\parbox[][7cm][t]{7cm}{
\caption{A plot of the value of $\Lambda$ as a function of the virial error.
Each iteration is marked by a plus ``$+$'' and connected consecutively.  As
shown in the figure, the iterated models reach a point of closest approach to
the mathematical solution, and then slowly drift towards less accurate models.}
\label{fig:iterate}}
\end{tabular}
\end{figure}

In addition to the previous test, it is also possible to apply the virial
theorem to obtain a measure of the accuracy of the model's structure. In what
follows we use the following formulation of the theorem:
\begin{equation}
0 = \int_V \rho_o \Omega_\star^2 r^2 \sin^2\theta dV
+ \frac{1}{2} \int_V \rho_o \Psi_o dV
+ \frac{3}{N+1} \int_V P_o dV,
\label{eq:virial}
\end{equation}
where $\rho_o = H^N$, $P_o = H^{N+1}$ and 
$\Psi_o = \psi_o/h_c$.
For a sufficient number of iterations, it is possible to attain a precision of
$10^{-13}$ on the virial test.  Beyond this point, successive iterations are
useless and can actually decrease the accuracy of the model.  In \fig{iterate},
we follow the evolution of $\Lambda$ and the virial error with each iteration. 
As can be seen in the figure, there are two phases: a first phase in which the
model is approaching the mathematical solution to the problem, and a second
phase in which the maximum precision has been attained and the model is slowly
drifting towards less accurate solutions.  For some of the rotating
configurations and with a well adjusted resolution, this second phase does not
contain a slow drift but remains close to a fixed point.  Either way, the best
point at which to stop the iterative scheme is at the transition between the
two phases.

The models on which are based the pulsation frequencies do not attain as high a
precision, because the iterative scheme was stopped before the transition
between the two phases.  This is because we use a small parameter called
$\epsilon$ which controls the relative error on the enthalpy and serves as a
stopping criteria.  If the value of $\epsilon$ is too low, than the iterative
program never reaches this precision on the enthalpy and therefore does not
output the stellar model.  We therefore set $\epsilon = 10^{-8}$ in most
calculations, which ensures successful convergence but reduces the accuracy of
the model.  As a result, the virial test typically attains a precision of
$4 \times 10^{-10}$.  For the non-rotating model, $\alpha$ takes on a
value around $54.182473$, which starts differing at the $5^\mathrm{th}$ digit
after the decimal point from the value given in \citet{Seidov04} and
corresponds to a relative precision of $\sim 10^{-7}$.


\subsection{Comparison with \citet{S81}}

\citet{S81} gives second order perturbative calculations for polytropic models.
Based on his coefficients, it is possible to obtain pulsation frequencies
via the following formula:
\begin{equation}
\omega = \omega_0 - \left( 1-C_1 \right) m \Omega + \left\{
\left( X_1 + X_2 + Z \right) + m^2 \left( Y_1 + Y_2\right) \right\}
\frac{\Omega^2}{\omega_0}.
\end{equation}
In order to compare our results with his, it is necessary to extract
perturbative coefficients from our own results.  The procedure used to find
these coefficients is fully described in \sect{least.squares}.  Before applying
this procedure, we first had to express our results in the same units as
\citet{S81} via the following conversion rule:
\begin{eqnarray}
\omega_{S81} &=& \sqrt{3\alpha_{nr}} \omega, \\
\Omega_{S81} &=& \sqrt{\frac{3\alpha_{nr}}{\Lambda}} \Omega_{\star},
\end{eqnarray}
where the subscript ``nr'' means ``non-rotating'' (\textit{i.e.} the value of
the parameter for the non-rotating polytrope) and the subscript ``S81'' means
that the quantity is in Saio's units.  It turns out that in Saio's units, the
mass of the polytrope depends on the rotation rate.  A comparison between his
coefficients $\left( \omega_0^2, C_1, X_1+X_2+Z, Y_1+Y_2 \right)$ and ours
showed a qualitative agreement between the two (to within $2\,\%$).  This
reduces the possibility of programming errors affecting our results.

\subsection{Comparison with \citet{C84}}

\begin{figure}[ht]
\includegraphics[width=8.8cm]{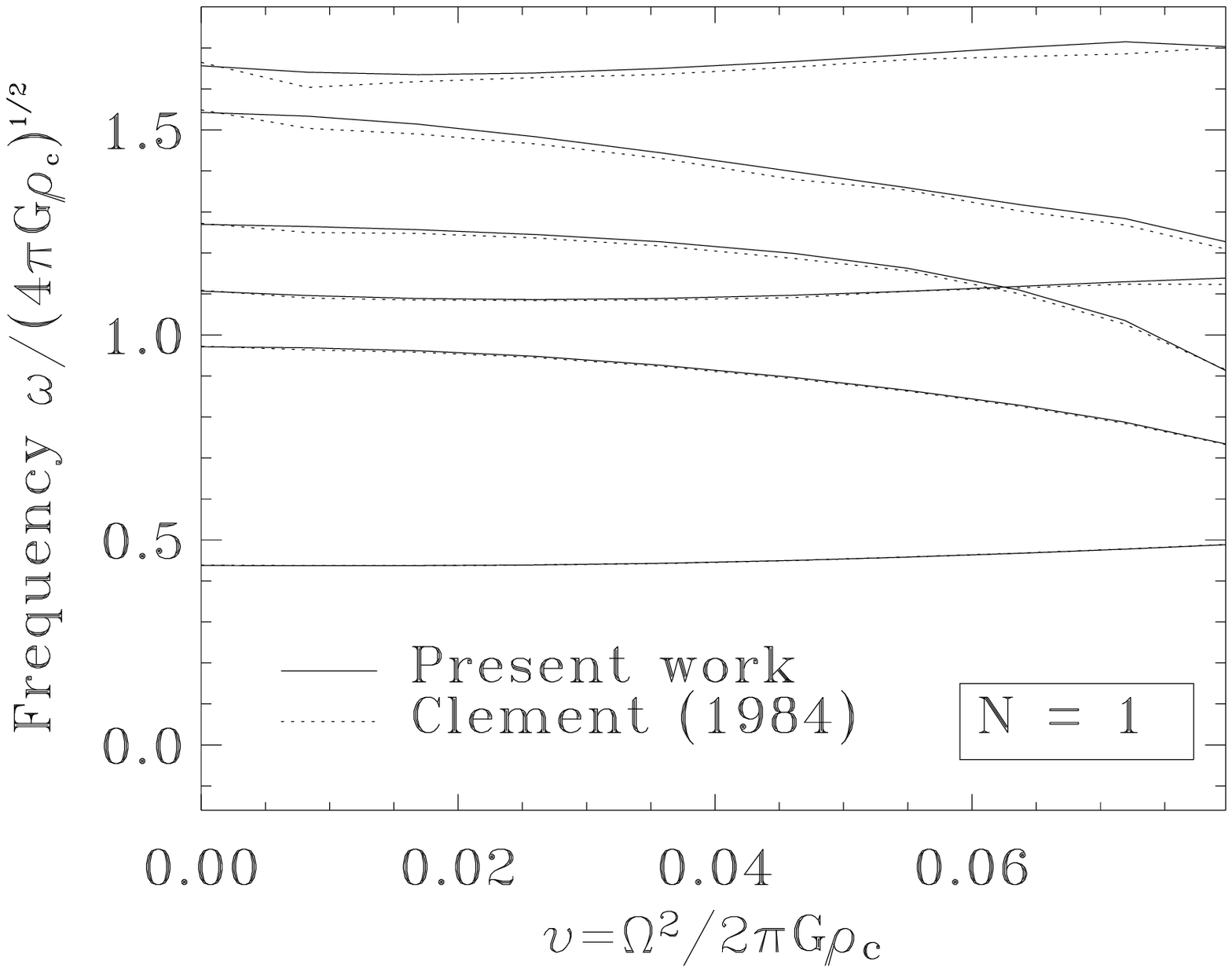}
\includegraphics[width=8.8cm]{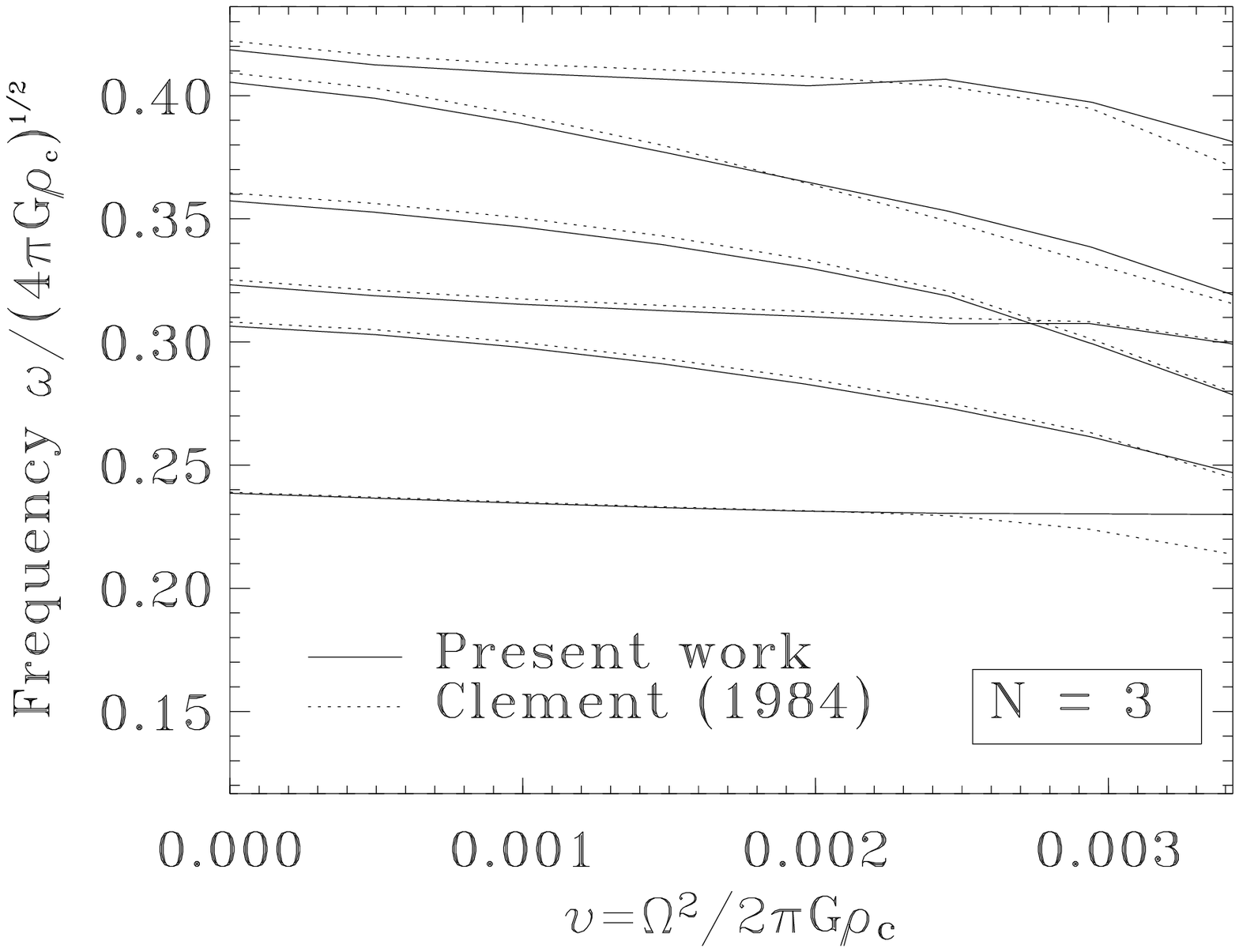}
\caption{A plot showing the frequencies in \citet{C84} and our calculations
of the same frequencies}
\label{fig:C84}
\end{figure}

Further comparisons can be done with \citet{C84}, who applied non-perturbative
techniques to calculate pulsation frequencies for $N=1$ and $N=3$ polytropes.
His frequencies are given in the same units as ours and no conversion is
needed. However, \citet{C84} used a rotational parameter which he called
``$\alpha$'' (denoted here as $\alpha_\mathrm{C84}$ so as to avoid confusion
with $\alpha$ from \eq{param}) and which is ``neither dimensionless nor
scale-free''.  Therefore, based on the conversion given in Tables~1 and~2 of
\citet{C84} and following his recommendations, we used the parameter $\upsilon
= \Omega^2/2\pi G\rho_c$ instead.  We allowed for uncertainties in the last
digit of $\upsilon$ and therefore calculated a corresponding range of
frequencies.  For example, if $\upsilon = 1.69\times 10^{-2}$ we would
calculate the frequencies corresponding to $\upsilon = 1.69\times 10^{-2} \pm
5\times 10^{-5}$.  The results are presented in \fig{C84}.

As can be seen from \fig{C84} the two sets of results qualitatively agree,
which once more makes numerical programming mistakes unlikely in our
calculations. However, Clement's results usually do not lie in the frequency
intervals we calculated (this would require a 4-digit accuracy).  This is
partially due to the fact that it is difficult to accurately reproduce the
polytropic models he used.  In order to illustrate this, we can use the
different parameters ($\upsilon, \,\rho_c/\bar{\rho}, \,\Req/\Rp,\, g_e/g_p$)
provided in Tables~1 and~2 of \citet{C84}, allow for uncertainties in the last
digit, and calculate the corresponding ranges for $\Omega_\star$.  For a given
rotation rate, if all the digits in the four parameters are accurate, then the
four different ranges for $\Omega_\star$ should overlap and give a more precise
idea as to the underlying model.  However, it was only possible to obtain at
most three overlapping ranges, and not four.  A typical example for the $N=1$
polytrope (with $\alpha_{C84} = 0.004$) is:
\begin{equation}
\begin{array}{*{7}l}
\upsilon &=& 3.57 \times 10^{-2} \pm 5 \times 10^{-5}
&\Rightarrow& \Omega_\star &=& 0.4615 \pm 4 \times 10^{-4}, \\
\rho_c/\bar{\rho} &=& 3.40 \pm 5 \times 10^{-3}
&\Rightarrow& \Omega_\star &=& 0.4580 \pm 9.4 \times 10^{-3}, \\
\Req/\Rp &=& 1.162 \pm 5 \times 10^{-4}
&\Rightarrow& \Omega_\star &=& 0.4590 \pm 7 \times 10^{-4}, \\
g_e/g_p  &=& 0.686 \pm 5 \times 10^{-4}
&\Rightarrow& \Omega_\star &=& 0.4621 \pm 4 \times 10^{-4}.
\end{array}
\end{equation}
This shows that the error bars we used on Clement's parameters are too small
and that the uncertainties on his models are larger.  Nonetheless, these
uncertainties do not fully account for the discrepancies between our
frequencies and his.  This can be shown by the fact that even for non-rotating
configurations (where there is no ambiguity on the underlying model) the
differences are still of the same order of magnitude.

In general, a quantitative comparison between our results and those of
\citet{S81} and \citet{C84} showed an agreement to $2$ or $3$ significant
digits. While providing a correct qualitative picture, the precision of these
studies is insufficient for future missions such as COROT.  COROT will observe
pulsation frequencies within the range of $0.1-10 \,\mathrm{mHz}$ with an
accuracy of $0.6\,\mu \mathrm{Hz}$ for the $20$ day runs and $0.08\,\mu
\mathrm{Hz}$ for the $150$ day runs \citep[\textit{e.g.}][]{Baglin02}, meaning
that an accuracy of $3$ to $5$ digits is required.  Therefore, it is important
to show that our results meet up to this requirement through other more
constraining tests and comparisons.

\subsection{Comparison with \citet{CDM94}}

\citet{CDM94} give very accurate frequencies for several \emph{non-rotating}
polytropic models.  Comparing our results with theirs provides a robust test
for accuracy.  In order to convert our frequencies $\omega$ into their units,
we apply the following conversion rule:
\begin{equation}
\nu_{CDM} = \nu_g \sqrt{3\alpha} \omega,
\end{equation}
where $\nu_\mathrm{CDM}$ is our frequency in their units, $\nu_g = 99.855377
\,\mu\mathrm{Hz}$, and $\alpha$ is given by \eq{param}.  A comparison between
their frequencies and ours revealed a very good agreement ($\Delta \omega/
\omega \sim 10^{-7}$ for a $N = 3$ polytrope and $\Delta \omega/\omega \sim
10^{-8}$ for a $N = 1.5$ polytrope at $\Omega=0$).  The modes which were
compared are: $\l = 0$ to $3$, $n = 1$ to $10$ for $N=3$ and $n = 15$ to $25$
for $N = 1.5$. The differences come from round-off errors in the last digit (if
we keep the same number of significant digits).

\subsection{Comparison with Paper~I}

We finally compared our results with those of  Paper~I.  In order to do
this comparison, it is necessary to remove the Coriolis force and to make the
Cowling approximation.  No conversion rule is necessary since both sets of
results are given in the same units.  The two sets of frequencies agree quite
well, even at large rotation rates ($\Delta \omega/\omega \sim 10^{-7}$).  This
result is significant due to the fact that the set of equations used in
Paper~I is entirely different than the one used here.

\subsection{Variational test}

The variational principle provides an integral formula which relates a
pulsation frequency to the structure of the corresponding mode 
\citep[\textit{e.g.}][]{LO67}.  It is therefore possible to apply this formula to a numerically
calculated eigenmode to obtain a ``variational frequency''.  The error on
this frequency is quadratic in the error of the eigenfunction 
\citep[\textit{c.f.}][]{CDM94}.  By comparing this frequency with the one obtained directly, it
is possible to estimate the accuracy of the results.  We used the following
non-dimensional formulation of the variational principle:
\begin{equation}
\label{eq:tests.var}
\omega^2 \displaystyle \int_V \rho_o |\vect{v}|^2 dV 
- |\omega|^2 \left( \int_V \frac{|p|^2 dV}{\rho_o c_o^2} 
- \int_{V_{\infty}} \left| \grad \Psi \right|^2 dV \right)
+2 i \omega \int_V \rho_o 
\vect{\Omega} \cdot \left( \vect{v}^* \times \vect{v} \right) dV 
- \int_V \rho_o N_o^2 \left| \vect{v} \cdot \vect{e}_g \right|^2 dV = 0,
\end{equation}
where $V$ is the volume of the star, $V_{\infty}$ is infinite space, and
$\vect{e}_g$ the unit vector in the same direction as the gravity vector. The
pressure is replaced by $H^{N}\Pi$ and the different integrals are performed
numerically\footnote{The numerical integration was based on Gauss' quadrature
method and a spectral expansion, using a radial resolution of $101$ points and
an angular resolution of
$200$ points.} which gives a second degree equation in $\omega$.  Solving this
equation gives the variational frequency which can then be compared with the
direct calculation of $\omega$.  Generally, we find differences $\Delta
\omega/\omega \sim  10^{-8}$ or better between the two.  This can be compared
with the results of \citet{IL90} who found differences of $10^{-3}$ when they
applied the variational principle to their calculations.  An explicit
formulation of the variational principle in spheroidal geometry is given in
\app{variational}.

\subsection{Influence of the parameters from the numerical method}

A final test consists in modifying different input parameters and seeing the
effect it has on the results.  We have therefore applied this test to a few
modes which are representative of all the modes that have been calculated.  The
parameters that were modified are the radial resolution $\Nr$ (which is the
same for the equilibrium model and the pulsation mode), the harmonic resolution
of the model $\Lmod$, the harmonic resolution of the pulsation mode $\Lmax$,
the shift $\sigma$ used in the Arnoldi-Chebyshev algorithm\footnote{The shift
comes from shift-and-invert methods and corresponds to a trial value around
which the Arnoldi-Chebyshev algorithm looks for frequencies. See
\citet{Valdetarro06} for an extensive discussion on the role of the shift in
numerical errors.}, and $\epsilon$ which controls the relative error of the
enthalpy in the equilibrium model. \Tab{stability.modes} lists the values used
for the different parameters and the induced frequency variations.  For a given
parameter, we used the frequency obtained at highest resolution (or lowest
value for $\epsilon$) as a reference.  In most cases, we obtained a rough
plateau at the different levels given in \tab{stability.modes}.  In some cases
however, there was a definite decrease of the error.  For instance, for those
modes in which it was tested, the error was roughly proportional to
$\epsilon$.  Also, for high frequency modes, the error strongly decreased as
$\Nr$ increased, as could be expected for high radial orders.  In the table, we
put the lower/final values of the error for both of these parameters.  The
information on the shift is slightly different.  The line ``Values'' gives the
amplitude of the variation on the value of the shift.  The next two lines
contain the standard deviation of the results.

The results on $\epsilon$ are not representative of the calculated
frequencies.  As was pointed out in \sect{model.precision}, the number
iterations was usually less than optimal because of a large value of $\epsilon$
($10^{-8}$ instead of $10^{-10}$), thus resulting in a decreased accuracy.  The
relative error on low frequency modes is $10^{-8}$ and that of high frequency
modes $10^{-7}$.

\begin{table}[ht]
\caption{Frequencies variations in terms of different parameters. $\Lmax$ is
the harmonic resolution of the pulsation modes, $\Lmod$ the harmonic resolution
of the models and $\Nr$ the radial resolution. The line ``Values'' gives the
different values that were used for the resolutions, $\epsilon$ and the width
giving the variation of the shift. The two following lines give the order of
magnitude of the induced frequency variations (in units of $\sqrt{4\pi G
\rho_c}$).}
\label{tab:stability.modes}
\begin{tabular}{*{6}l}
\hline
\hline
 & $\Lmax$ & $\Lmod$ & $\Nr$ & Shift & $\epsilon$ \\
\hline
Values & $40$, $44$ ... $80$ & $30$, $34$ ... $70$ & $32$, $36$ ... $60$ &
$2-5 \times 10^{-4}$ & $10^{-8}$...$10^{-10}$ \\
Low frequency modes & $< 10^{-15}$ & $10^{-10}$ & $10^{-10}$ & $10^{-13}$ &
$10^{-10}$\\
High frequency modes & $10^{-14}$ & $10^{-9}$ & $10^{-10}$ & $10^{-11}$ & $10^{-9}$ \\
\hline
\end{tabular}
\end{table}

\subsection{Discussion}

Overall, the main source of error in the present calculations is the
uncertainties on the equilibrium model.  This is because we chose a convergence
criteria which was sure to be met, but which lead to a number of iterations
less than optimal.  This therefore leads to a global accuracy of  $7$ digits
after the decimal point (in units of $\sqrt{4\pi G \rho_c}$), the last digit
being uncertain.  \Tab{stability.modes} however shows that these calculations
could potentially be made more accurate.  The present accuracy is nonetheless
largely sufficient for the requirements of COROT, which will be at most $5$
significant digits.

\section{Results}

We now proceed to present the results themselves.  We followed acoustic
adiabatic pulsation modes (with $\Gamma_1 = 5/3$) from a zero rotation rate to
$0.59 \Omega_K$ (where $\Omega_K = \sqrt{GM/\Req^3}$ is the Keplerian break-up
rotation rate), using the same procedure as in Paper~I.  This involves
identifying the frequencies at $\Omega=0$, following their evolution while
progressively increasing the rotation rate, and working through a number of
avoided crossings.  The underlying polytropic models have an index $N=3$ which
gives a polytropic exponent $\gamma = 4/3$.  The modes that were calculated
are: $\l=0$ to $3$, $n=1$ to $10$ and $m=-\l$ to $\l$ both with and without the
Coriolis force.


\subsection{Comparison with perturbative methods}

In this section, we compare complete and perturbative calculations so as to
determine the range of validity of perturbative methods.

\subsubsection{Perturbative coefficients}
\label{sect:least.squares}

In order to compare perturbative calculations with complete ones, it proved
necessary to compute our own perturbative coefficients, since we were unable to
find perturbative coefficients for polytropic models with a sufficient accuracy
in the literature.  Instead of using the traditional method of perturbing the
fluid equations and finding corrections of various orders on the frequencies
(see \citet{SGD98} for a complete description),  we did a series of complete
calculations for small rotation rates ($\left(\Omega_\star\right)_i = 0$,
$10^{-6}$, $10^{-5}$, $10^{-4}$, $10^{-3}$, $0.002$, $0.004$ ... $0.018$) and
applied a least squares fit to the results.  In order to increase the accuracy
of such calculations, we made use of \eq{coeff.sym} and separated even and odd
powers of the rotation rate:
\begin{eqnarray}
\frac{\omega_{n,\,l,\,m} + \omega_{n,\,l,\,-m}}{2} &=& \omega_{n,\,l,\,m}^0 + 
                                                       \omega_{n,\,l,\,m}^2 \Omega^2 +
                                                       \omega_{n,\,l,\,m}^4 \Omega^4 +
                                                       \O(\Omega^6), \\
\frac{\omega_{n,\,l,\,m} - \omega_{n,\,l,\,-m}}{2} &=& \omega_{n,\,l,\,m}^1 \Omega + 
                                                       \omega_{n,\,l,\,m}^3 \Omega^3 +
                                                       \omega_{n,\,l,\,m}^5 \Omega^5 +
                                                       \O(\Omega^7).
\end{eqnarray}
By fitting $(\omega_{n,\,l,\,m} + \omega_{n,\,l,\,-m})/2$ and
$(\omega_{n,\,l,\,m} - \omega_{n,\,l,\,-m})/2$ the number of unknowns is
reduced to three and the residues are smaller.  It is necessary to include the
fourth and fifth powers of the rotation rate so as to ensure that the second
and third order coefficients are reasonably accurate.  The results are given in
\tab{perturbative.coefficients} for frequencies and rotation rates in units
of $\left( GM/\Rp^3 \right)$. From these coefficients the frequencies are
given through the following formula:
\begin{equation}
\label{eq:perturbative}
\omega = \omega_0 - m(1 - C)\Omega + \left(D_1 + m^2 D_2\right) \Omega^2
+ m\left(T_1 + m^2 T_2\right) \Omega^3 + \O\left(\Omega^4\right).
\end{equation}
The form of the second degree coefficients was obtained from \citet{S81} and
that of the third degree coefficients from Goupil (private communication).  In
order to express these results in units of $\Omega_K$ instead, one can use the
following perturbative formula:
\begin{equation}
\left( \frac{\Omega}{\Omega_K} \right) = 
\left( \frac{\Omega}{\Omega_K^\mathrm{pol}} \right)  +
A\left( \frac{\Omega}{\Omega_K^\mathrm{pol}} \right)^3  +
\O\left\{\left( \frac{\Omega}{\Omega_K^\mathrm{pol}} \right)^5 \right\},
\end{equation}
where $\Omega_K^\mathrm{pol} = \left( GM/\Rp^3 \right)$ and $A \simeq 0.77166$.

\begin{table}[h!]
\caption{Perturbative coefficients for a $N=3$ polytrope, deduced from complete
calculations.  The frequencies and the rotation rate are expressed in units of
$\left(GM/\Rp^3\right)^{1/2}$.}
\label{tab:perturbative.coefficients}
\begin{minipage}{0.44\linewidth}
\begin{tabular}{r@{.}lcccc}
\hline
\hline
\multicolumn{2}{c}{$\omega_0$} & $C$ & $D_1$ & $D_2$ & $T_1$ \\
\hline
\multicolumn{6}{c}{$\l = 0$} \\
\hline
 3&042155 & ... & -1.194 & ... & ... \\
 4&121230 & ... & -2.315 & ... & ... \\
 5&336900 & ... & -3.439 & ... & ... \\
 6&591212 & ... & -4.484 & ... & ... \\
 7&855027 & ... & -5.484 & ... & ... \\
 9&120432 & ... & -6.459 & ... & ... \\
10&384948 & ... & -7.416 & ... & ... \\
11&647767 & ... & -8.361 & ... & ... \\
12&908679 & ... & -9.298 & ... & ... \\
14&167704 & ... &-10.228 & ... & ... \\
\hline
\multicolumn{6}{c}{$\l = 1$} \\
\hline
 3&377036 & 0.0295367 & -1.072 & -1.030 & -0.04612 \\
 4&642432 & 0.0342809 & -1.760 & -1.699 & -0.04204 \\
 5&909240 & 0.0335303 & -2.402 & -2.315 & -0.02715 \\
 7&176668 & 0.0305143 & -3.019 & -2.901 & -0.01634 \\
 8&443277 & 0.0270732 & -3.621 & -3.470 & -0.00951 \\
 9&708372 & 0.0238467 & -4.213 & -4.028 & -0.00524 \\
10&971700 & 0.0210064 & -4.797 & -4.578 & -0.00254 \\
12&233222 & 0.0185621 & -5.375 & -5.122 & -0.00080 \\
13&492998 & 0.0164731 & -5.950 & -5.662 &  0.00034 \\
14&751133 & 0.0146882 & -6.521 & -6.198 &  0.00108 \\
\hline
\end{tabular}
\end{minipage} \hfill
\begin{minipage}{0.52\linewidth}
\begin{tabular}{r@{.}lccccc}
\hline
\hline
\multicolumn{2}{c}{$\omega_0$} & $C$ & $D_1$ & $D_2$ & $T_1$ & $T_2$ \\
\hline
\multicolumn{7}{c}{$\l = 2$} \\
\hline
 3&906874 & 0.1538359 & -1.578 & -0.294 &  0.02344 & 0.00527 \\
 5&169469 & 0.0818188 & -2.396 & -0.459 &  0.00493 & 0.00477 \\
 6&439990 & 0.0544285 & -3.146 & -0.606 &  0.00020 & 0.00307 \\
 7&708951 & 0.0403695 & -3.867 & -0.745 & -0.00087 & 0.00218 \\
 8&975891 & 0.0318248 & -4.572 & -0.880 & -0.00094 & 0.00169 \\
10&240946 & 0.0260651 & -5.267 & -1.013 & -0.00074 & 0.00139 \\
11&504260 & 0.0219090 & -5.955 & -1.144 & -0.00049 & 0.00119 \\
12&765953 & 0.0187647 & -6.636 & -1.273 & -0.00025 & 0.00105 \\
14&026134 & 0.0163027 & -7.314 & -1.402 & -0.00006 & 0.00094 \\
15&284901 & 0.0143246 & -7.987 & -1.530 &  0.00010 & 0.00086 \\
\hline
\multicolumn{7}{c}{$\l = 3$} \\
\hline
 4&294602 & 0.1193654 & -1.898 & -0.169 & -0.01155 & 0.00041 \\
 5&591067 & 0.0742468 & -2.728 & -0.240 & -0.00113 & 0.00157 \\
 6&878680 & 0.0517251 & -3.496 & -0.307 &  0.00015 & 0.00140 \\
 8&158826 & 0.0387755 & -4.236 & -0.371 &  0.00038 & 0.00113 \\
 9&433911 & 0.0305232 & -4.960 & -0.434 &  0.00043 & 0.00091 \\
10&705348 & 0.0248710 & -5.674 & -0.496 &  0.00044 & 0.00076 \\
11&973956 & 0.0207887 & -6.380 & -0.557 &  0.00046 & 0.00064 \\
13&240238 & 0.0177183 & -7.081 & -0.618 &  0.00048 & 0.00055 \\
14&504529 & 0.0153345 & -7.777 & -0.678 &  0.00050 & 0.00048 \\
15&767068 & 0.0134359 & -8.470 & -0.738 &  0.00051 & 0.00042 \\
\hline
\end{tabular}
\end{minipage}
\end{table}

In order to estimate the accuracy of these perturbative coefficients, there are
a number of tests that can be done.  First of all, the zeroth order
coefficients are simply the pulsation frequencies without rotation but are
treated as unknowns in the least squares development.  The frequencies without
rotation are recovered in the least squares fit to an accuracy of at least
$5.4\times 10^{-8}$ in the units of \tab{perturbative.coefficients}. The first
order coefficient $C$, can be calculated via integrals based on the zeroth
order solution \citep{Ledoux1951}.  This alternate way of calculating the
coefficients agrees to within $1.4\times 10^{-9}$ (this does not necessarily
mean that the coefficients are accurate to that precision but does show a high
degree of internal coherence).   For the second and third degree coefficients,
we checked to see if they satisfied the forms given in \eq{perturbative}; the
number of significant digits in \tab{perturbative.coefficients} has been
adjusted accordingly.  These forms were a constraint only on the $\l=2$ and $3$
second order coefficients and on the $\l=3$ third order coefficients.  Another
test we did consisted in applying the least squares fit to a subset of the
results used in the first fit and seeing whether the coefficients were
altered.  This test indicates roughly the same accuracy as the other tests.

\subsubsection{Comparison}

\begin{figure}[t!]
\begin{tabular}{ll}
\parbox[][19cm][t]{12cm}{
\includegraphics[width=12cm]{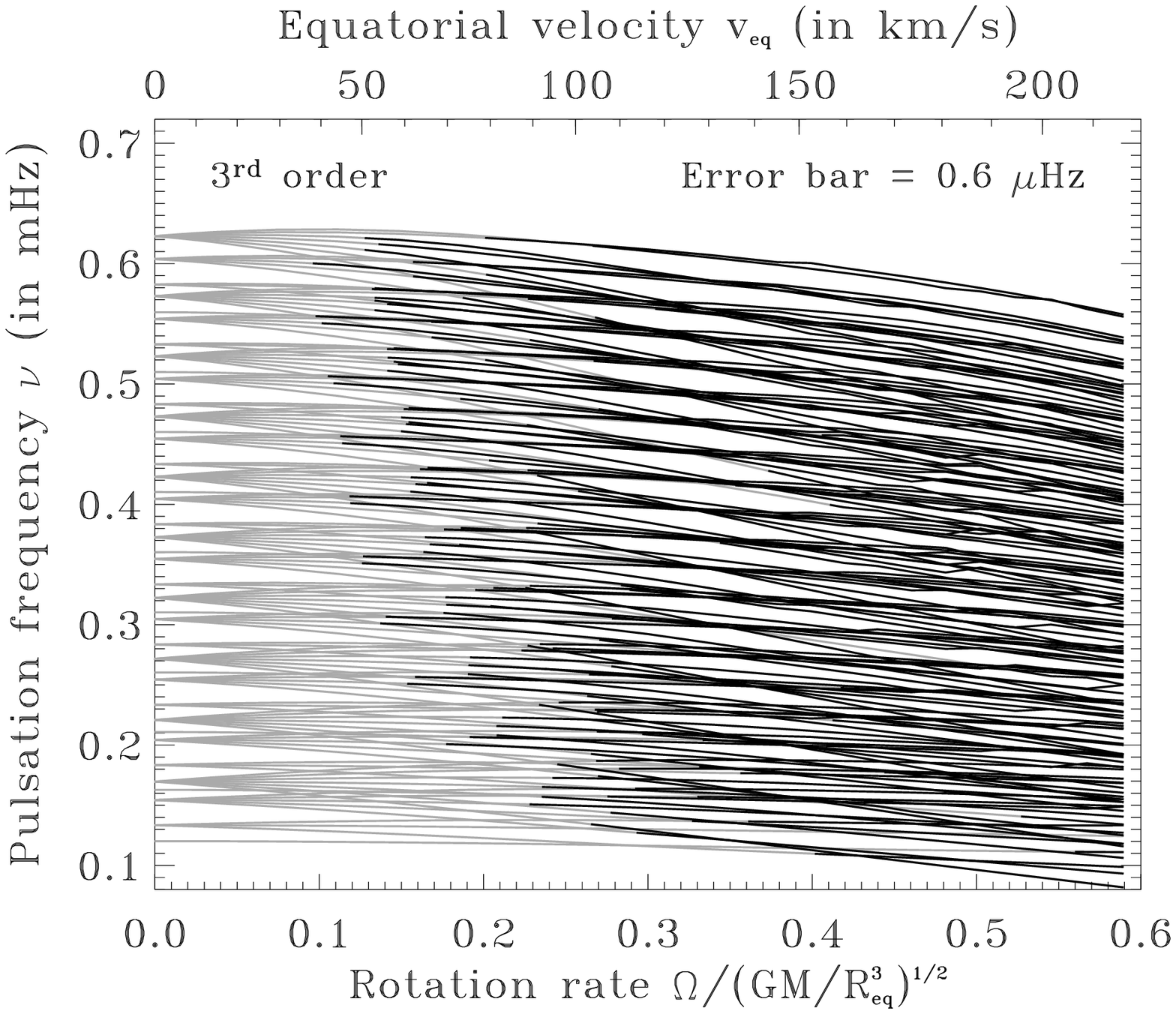} \\
\includegraphics[width=12cm]{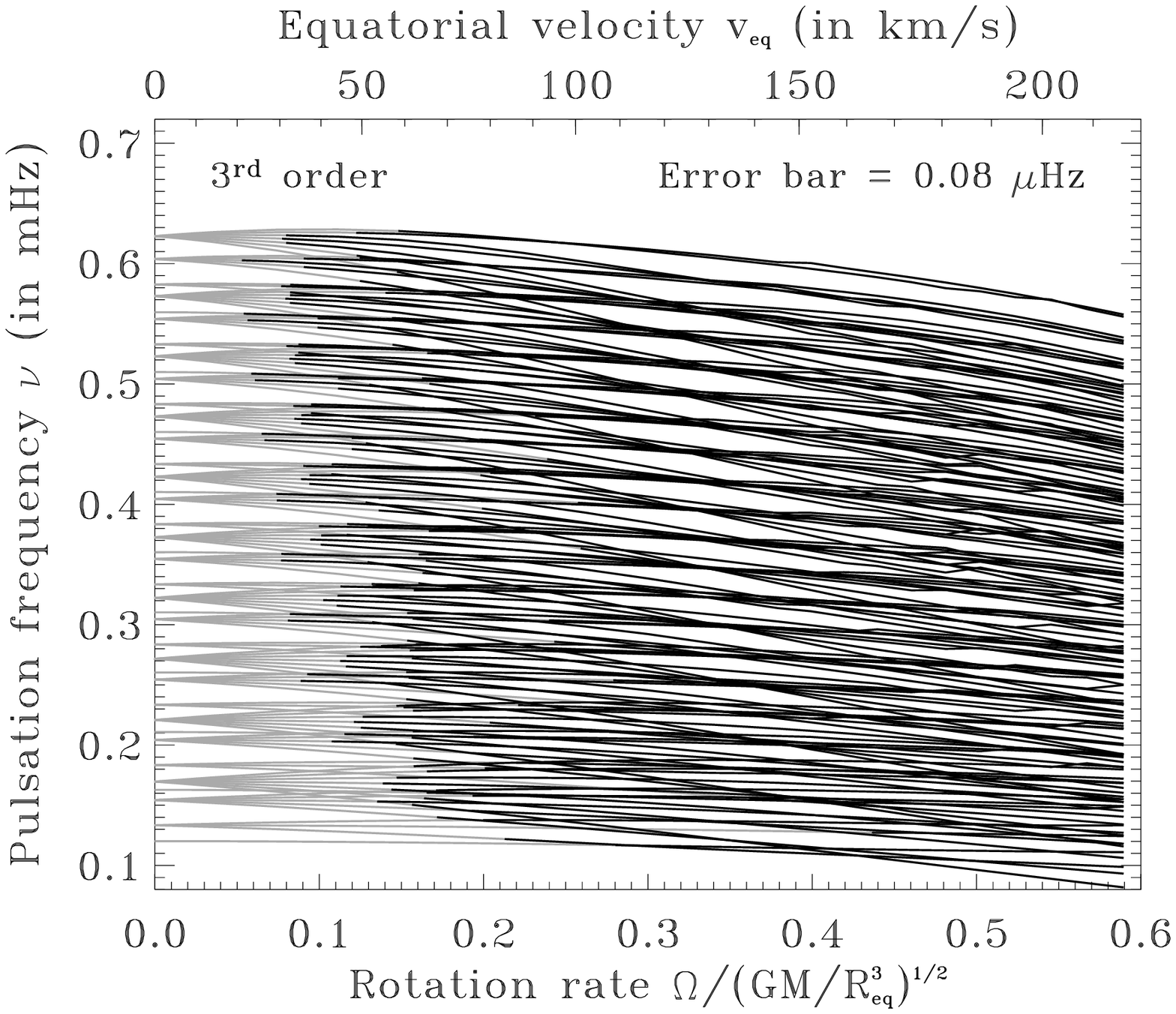} } &
\parbox[][19cm][t]{5.5cm}{\vspace*{10mm}
\caption{Plots of the evolution of pulsation cyclic frequencies ($\nu = \omega
/2\pi$) as a function of the rotation rate.  The frequencies are followed from
the non-rotating case to $0.59\,\Omega_K$ using the procedure described in
Paper~I.  At small rotation rates, it is easy to recognise the usual multiplet
structure as predicted from perturbative methods.  Once the rotation rate is
sufficient large, the multiplets are less regular and overlap, which greatly
complicates the interpretation of the oscillation spectrum.  Superimposed is
the domain of validity of $3^\mathrm{rd}$ order perturbative methods using
COROT's error bars of $0.6 \, \mu\mathrm{Hz}$ (upper panel) and $0.08 \, \mu
\mathrm{Hz}$ (lower panel).  The calculations were done with $N = 3$ polytropic
models with $\Rp = 2.3 R_\odot$ and $M = 1.9 M_\odot$.}
\label{fig:comparison}}
\end{tabular}
\end{figure}

Based on these coefficients, it is possible to calculate perturbative
frequencies which can then be compared to the complete calculations, thereby
establishing a domain of validity for perturbative methods. In
\fig{comparison}, we show two such domains for $3^\mathrm{rd}$ order methods,
one for each of COROT's error bars ($0.6 \, \mu \mathrm{Hz}$ for the 20 runs
and $0.08 \, \mu \mathrm{Hz}$ for the 150 day runs).  The underlying polytropic
models have a fixed mass of $1.9 \, M_\odot$ and a fixed polar radius of $2.3
\, R_\odot$,  both of which are typical of $\delta$ Scuti pulsators.  When the
distance between the perturbative frequency and the complete one exceeds
COROT's error bars, the frequency is shown in black. Otherwise, it is shown in
grey.  From these figures, it is clear that complete methods are required
beyond $\vsini = 75\,\kms$ for COROT's 20 day programs and $\vsini = 50\,\kms$
for COROT's 150 day programs.

It is important to bear in mind that the domain of validity obtained for
perturbative methods depends on the choice of rotational variable used in the
development.  In order to illustrate this, suppose we develop a frequency in
terms of two different rotational parameters $X$ and $Y$: $\omega = a_0 + a_1 X
+ a_2 X^2 + a_3 X^3 + \O(X^4) = b_0 + b_1 Y + a_2 Y^2 + a_3 Y^3 + \O(Y^4)$. 
When the relationship between $X$ and $Y$ is more complex than a simple
proportionality, the neglected terms, $\O(X^4)$ and $\O(Y^4)$, are not the
same. As a result, a $3^\mathrm{rd}$ order development in terms of $X$ or $Y$
will give different values for $\omega$, thus modifying the corresponding
domain of validity.  Therefore, we decided to compute the domain of validity
associated with the variables $\Omega/\Omega_K$ and $\Omega/(GM/\Rp^3)^{1/2}$
to see if there was a substantial difference between the two.  For individual
frequencies, there can be large differences, but when all the frequencies are
considered, the global result is roughly the same.

\begin{figure}[ht!]
\begin{tabular}{ll}
\parbox[][9cm][t]{12cm}
{\includegraphics[width=12cm]{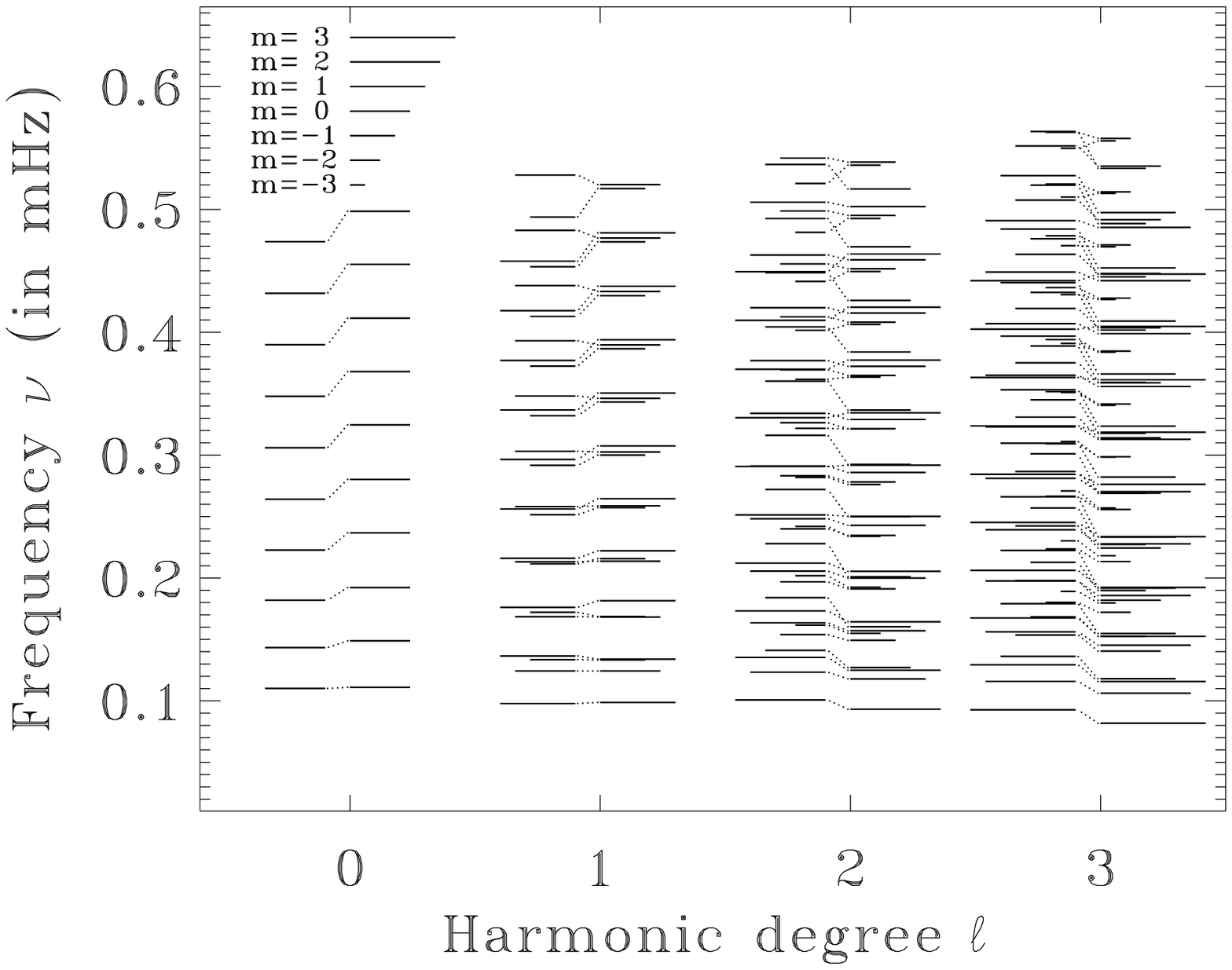}} &
\parbox[][9cm][t]{5.5cm}
{\caption{This figure shows a comparison between perturbative cyclic
frequencies and complete ones at $0.59 \Omega_K$.  The equilibrium model is the
same as in \fig{comparison}.  The complete calculations are represented by the
bars that go to the right and the perturbative ones by the bars going to the
left. The length of the bars gives the azimuthal order.  Complete and
perturbative calculations are connected by dotted lines (the correspondence is
based on mode labelling as described in Paper~I).  It is clear from this figure
that perturbative calculations lead to substantial error at high rotation
rates, and cannot correctly anticipate the order of the modes.  Even if the
perturbative frequencies were multiplied by a global corrective factor, the
agreement remains poor.}
\label{fig:differences}}
\end{tabular}
\end{figure}

In \fig{differences}, we show the differences between complete frequencies and
perturbative ones at $0.59 \Omega_K$.  We have kept the same parameters for the
equilibrium model as in \fig{comparison}.  As can be seen from the figure,
differences between the two sets of calculations are substantial and comparable
to the large frequency separation (which seems to survive rotation).  The order
of frequencies is not the same between the two sets of calculations.  As a
result, it is necessary to use complete calculations in order to correctly
interpret a pulsating star rotating at such a high rotation rate. 

Recently, \citet{Suarez05} attempted to model Altair through asteroseismology.
The effects of rotation were included in the pulsation modes using
$2^\mathrm{nd}$ order perturbative methods. Later interferometric studies
suggested an equatorial velocity of $280\,\kms$ \citep{Domiciano05}, which is
above $216 \, \kms$, the equatorial velocity corresponding to \fig{differences}
(if we use a mass of $M=1.8\,M_\odot$ and a polar radius $\Rp = 1.7\,R_\odot$
instead, we obtain $v_\mathrm{eq} = 244 \,\kms$).  As a result, it is pretty
obvious that what is required in \citet{Suarez05} is complete calculations of
the effects of rotation before being able to interpret Altair's oscillation
spectrum (not to mention complete models of rapidly rotating stars).

\subsubsection{Relative importance of the Coriolis and centrifugal forces}

\begin{figure}[ht]
\begin{tabular}{ll}
\includegraphics[width=8.7cm]{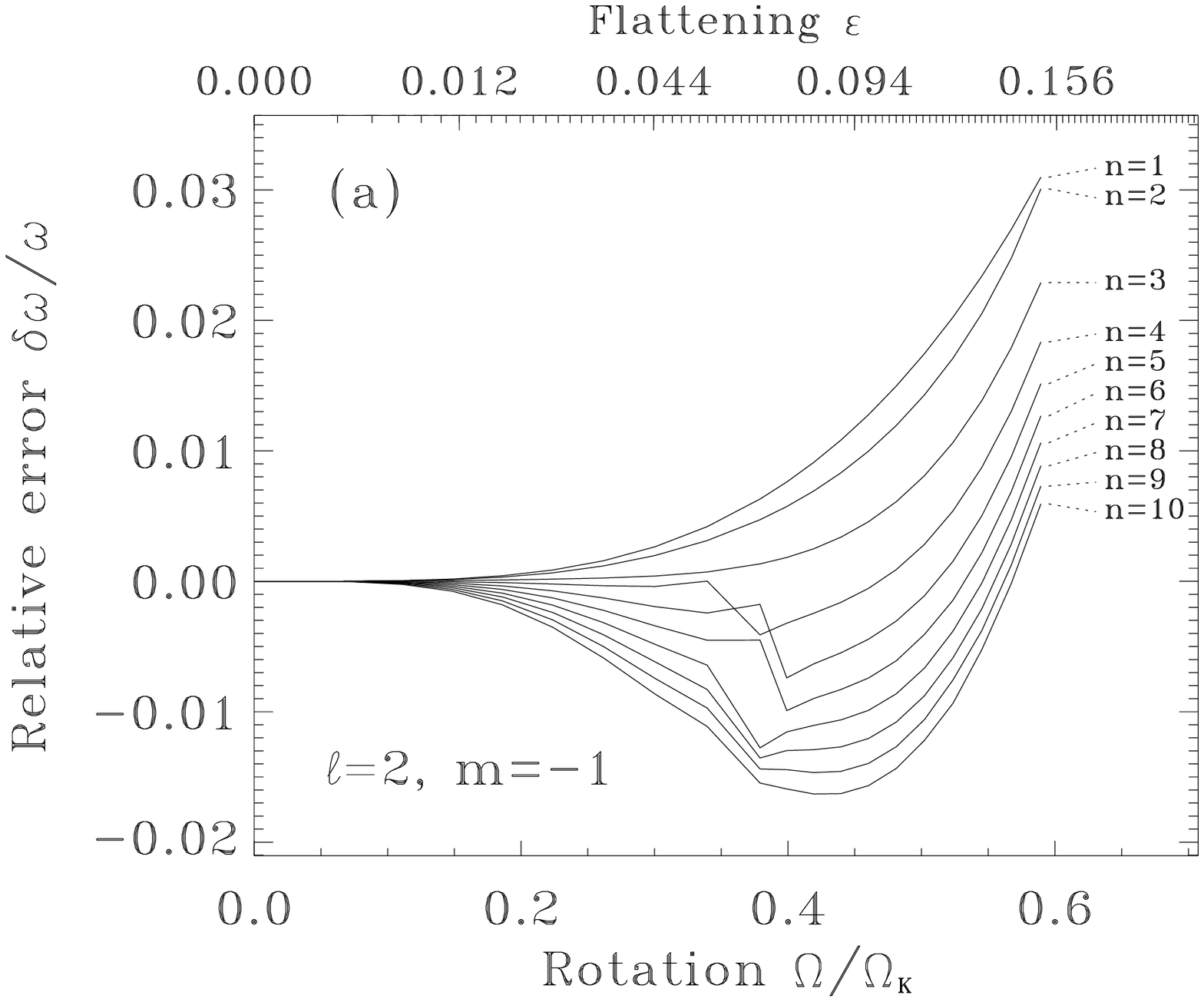} &
\includegraphics[width=8.7cm]{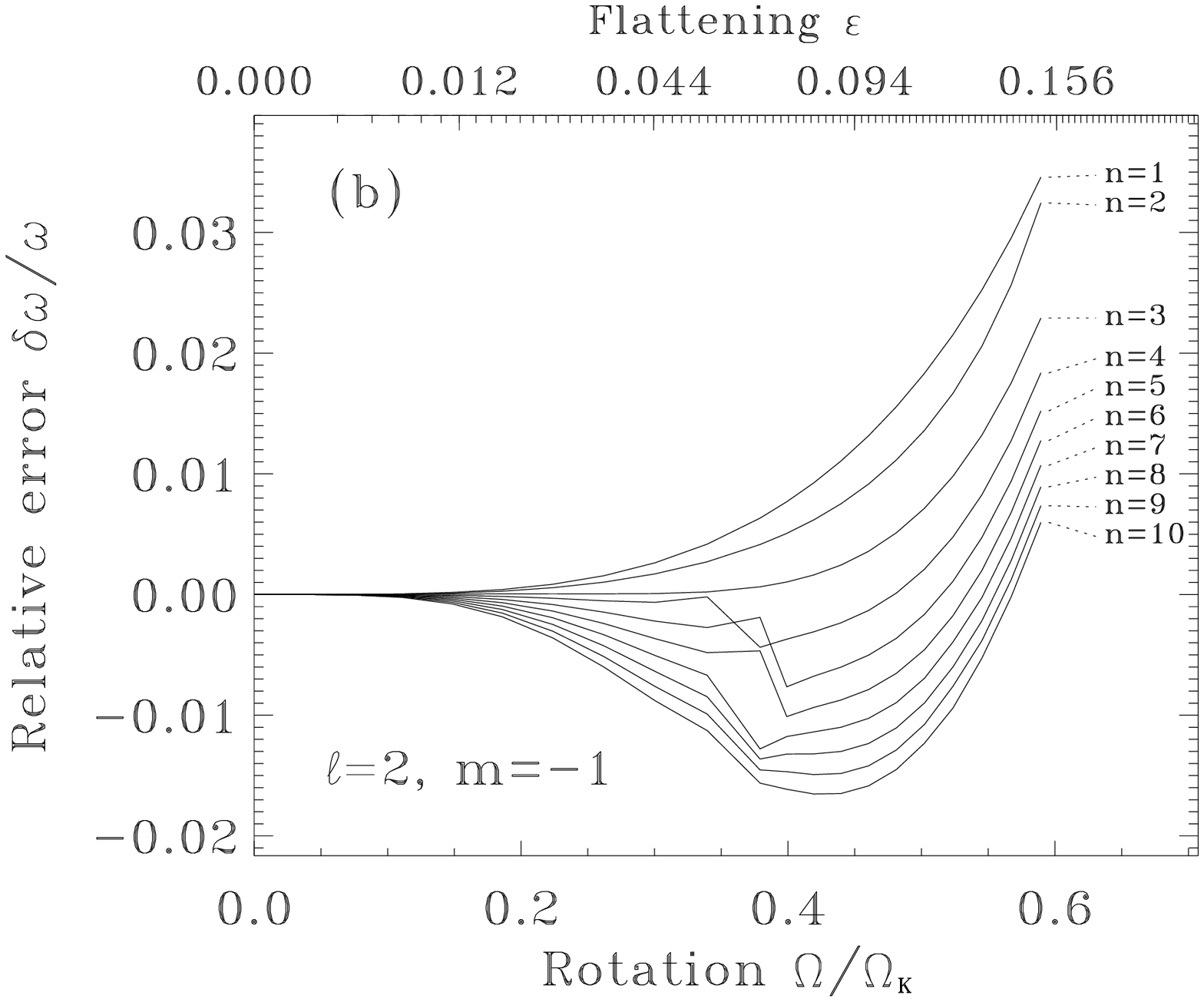} \\
\parbox[][6.85cm][t]{8.7cm}{\includegraphics[width=8.7cm]{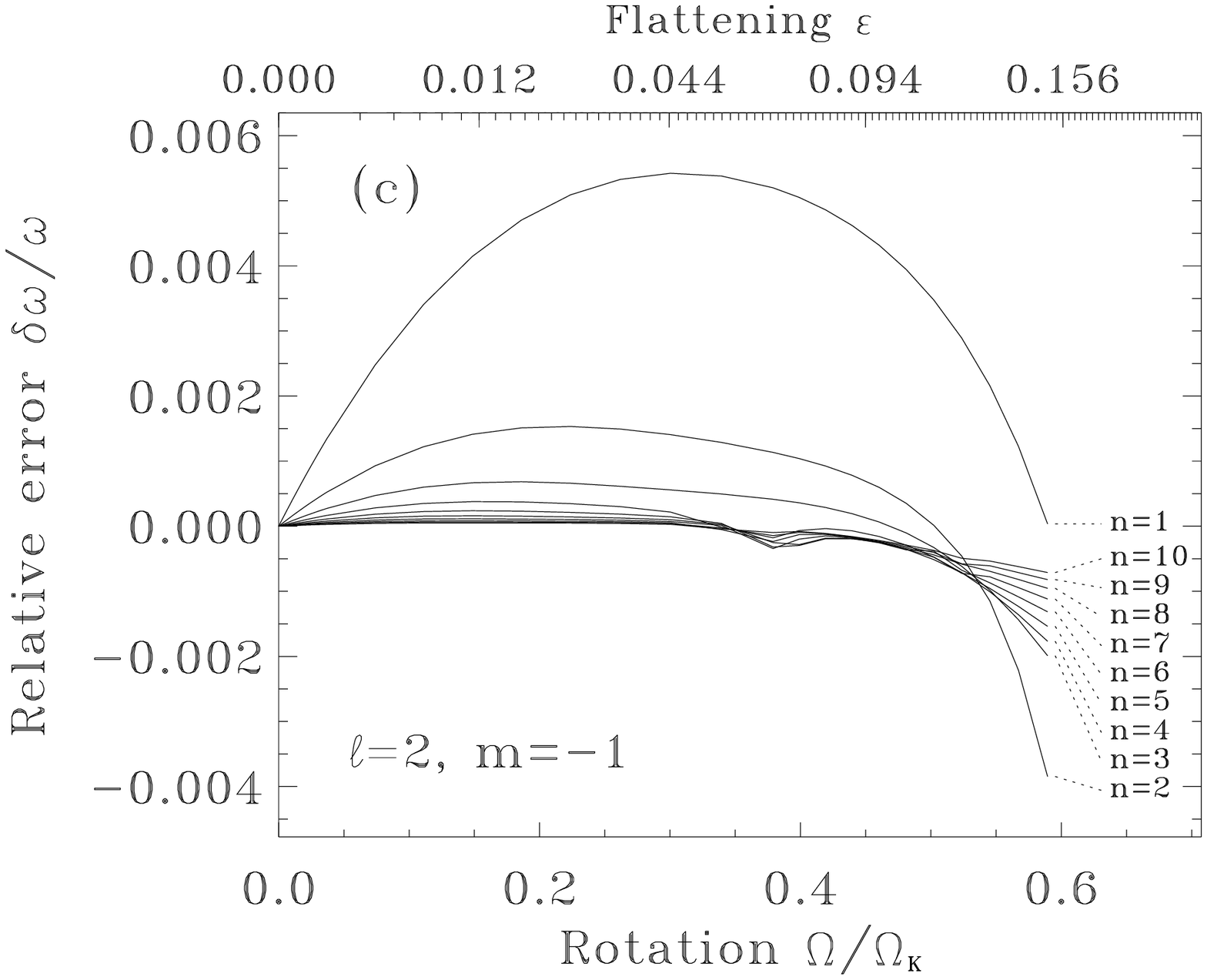}} &
\parbox[][6.85cm][t]{8.45cm}
{\vspace*{8mm}
\caption{Relative error of different calculations of $\l =2$, $m=-1$ modes of
various radial orders.  The radial orders are indicated on the right side of
each panel.  \textbf{Panel (a)}: the relative error from using $3^\mathrm{rd}$
order perturbative methods.  \textbf{Panel (b)}: Same as panel (a) except that
the Coriolis force has been excluded from both perturbative and complete
calculations.  \textbf{Panel (c)}:  the relative error which comes from
neglecting the Coriolis force.  Explicit expressions for the various errors are
given in \eqtoeq{error_a}{error_c}.  The similarity between panels (a) and (b)
and the differences with panel (c) show the dominant role of the centrifugal
force in errors related to perturbative methods.}
\label{fig:centrifugal}}
\end{tabular}
\end{figure}

It is then interesting to analyse what is the main source of differences
between perturbative calculations and complete ones.  In \fig{centrifugal} we
show three different graphs which give the relative errors associated with
different calculations:
\begin{eqnarray}
\label{eq:error_a}
\left( \frac{\delta \omega}{\omega} \right)_{(a)} &=& 
\frac{\omega_\mathrm{pert.} - \omega}{\omega}, \\
\left( \frac{\delta \omega}{\omega} \right)_{(b)} &=& 
\frac{\omega_\mathrm{pert.}^\mathrm{no\,\,Cor.} - \omega^\mathrm{no\,\,Cor.}}
{\omega^\mathrm{no\,\,Cor.}}, \\
\label{eq:error_c}
\left( \frac{\delta \omega}{\omega} \right)_{(c)} &=& 
\frac{\omega^\mathrm{no\,\,Cor.} - \omega}{\omega},
\end{eqnarray}
where the subscripts $(a)$, $(b)$ and $(c)$ correspond to the different panels
in \fig{centrifugal}, the subscript ``pert.'' to $3^\mathrm{rd}$ order
perturbative calculations and the superscript ``no Cor.'' to calculations done
without the Coriolis force.  From these panels, it is possible to deduce the
dominant role of the centrifugal force in the differences between perturbative
and complete calculations.  Panels~(a) and~(b) are very similar, yet the first
one includes the Coriolis force and the second one excludes it.  Panel~(c)
shows the errors which come from excluding the Coriolis force.  These errors
are at least ten times smaller than in cases~(a) and~(b) and decrease with the
radial order.  This decrease is expected because as the radial order $n$ of the
mode increases, the time scale of the oscillations decreases and becomes much
shorter than the $1/\Omega$ time scale associated with the Coriolis force.  As
a result high order modes are less affected by the Coriolis force.

\begin{figure}[ht]
\begin{tabular}{ll}
\parbox[][][t]{8.7cm}{\includegraphics[width=8.7cm]{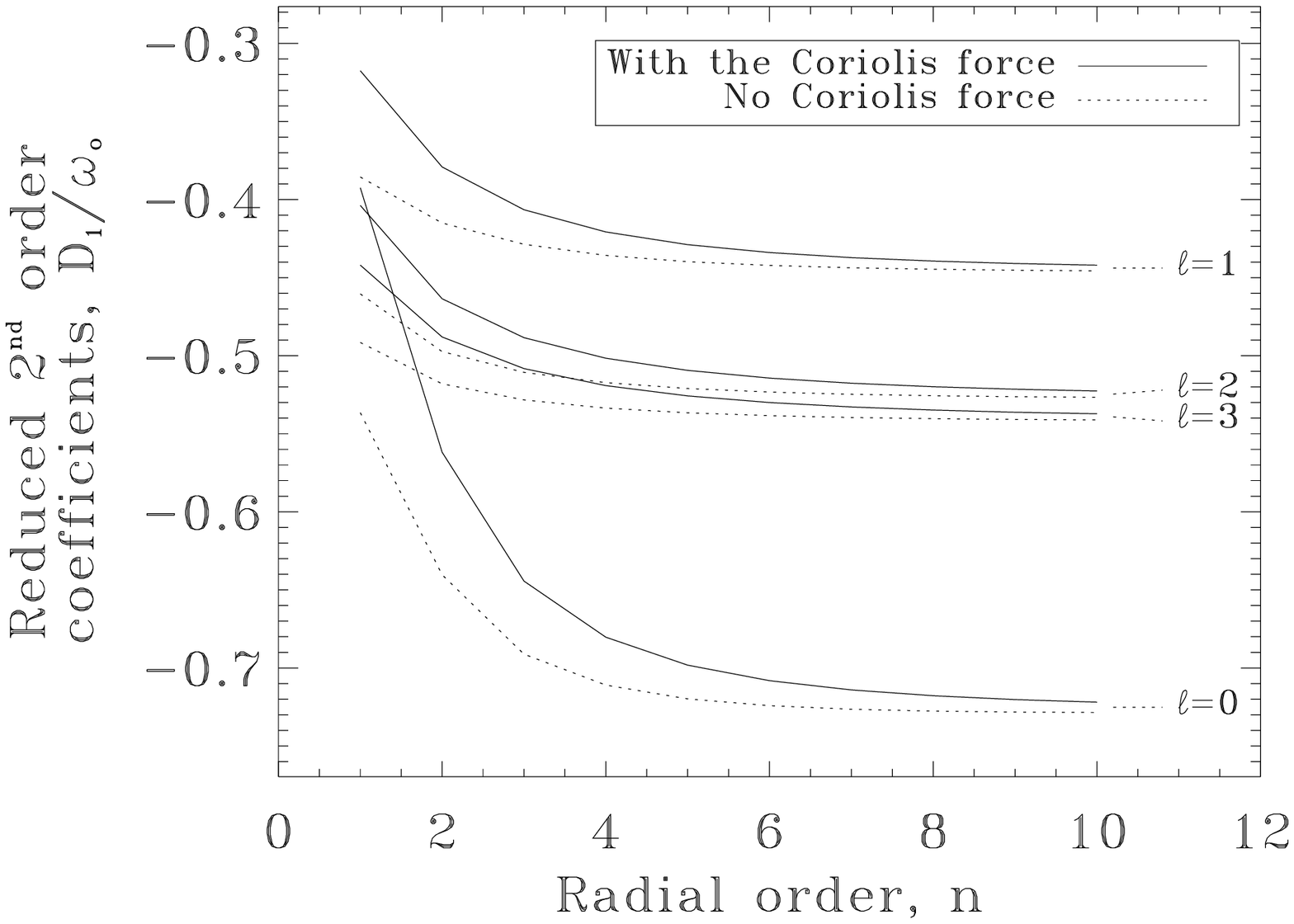}} &
\parbox[][][t]{8.7cm}{\includegraphics[width=8.7cm]{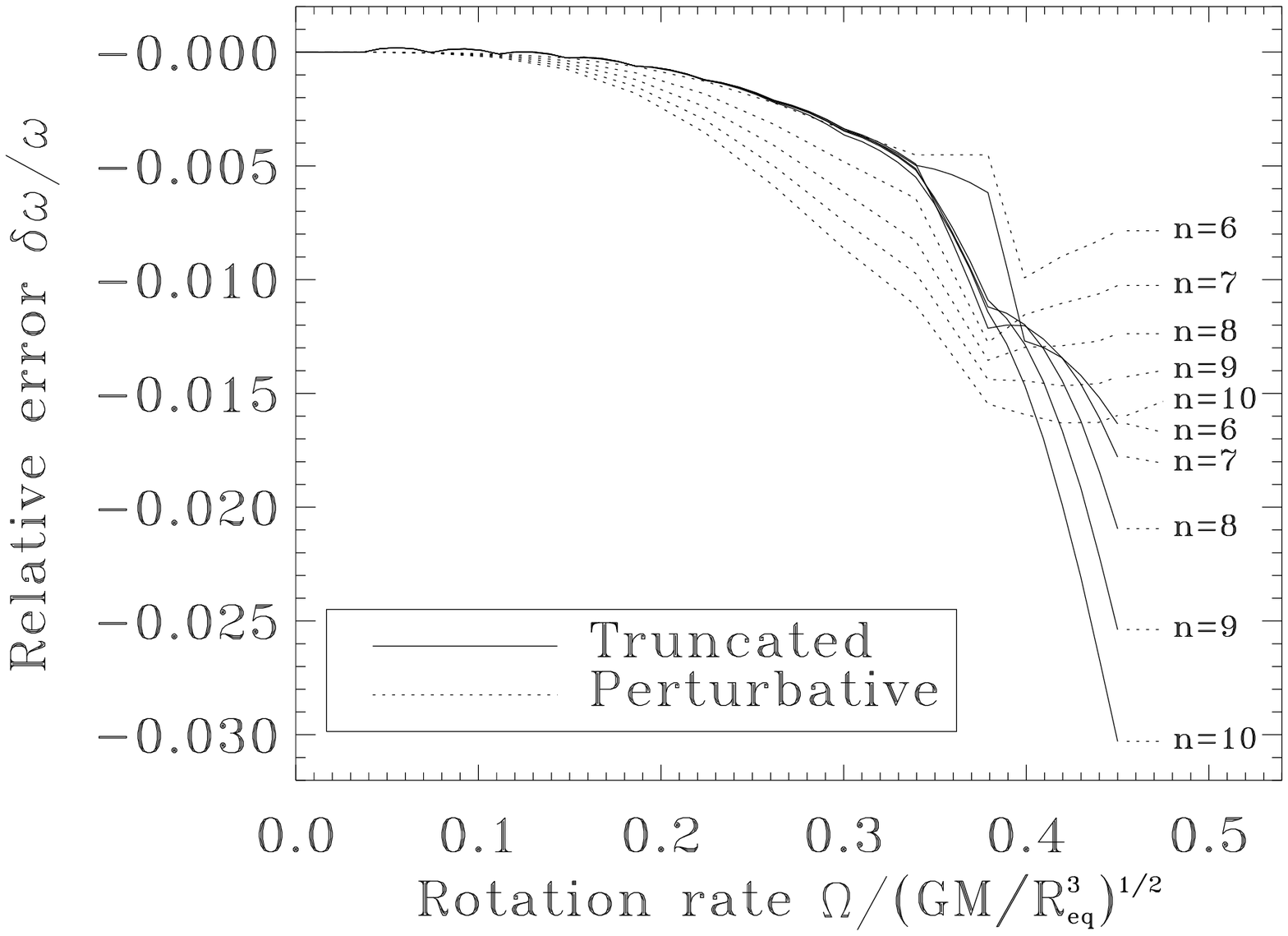}} \\
\parbox[][4.3cm][t]{8.45cm}
{\caption{Ratio of $2^\mathrm{nd}$ order perturbative coefficients to
corresponding frequencies at $\Omega=0$, as a function of the radial order $n$.
As $n$ increases, these ratios approach a constant value, which shows the
proportional effects of the centrifugal deformation of the star.}
\label{fig:coeff}} &
\parbox[][4.3cm][t]{8.45cm}
{\caption{A comparison between relative errors in perturbative calculations and
those in truncated calculations (in which the equilibrium model is reduced to
$2$ spherical harmonics and the pulsation mode to $2$ poloidal + $2$ toroidal
harmonics).  This figure shows that truncated calculations are closer to
complete ones than perturbative calculations, which shows the importance of
retaining higher powers of $\Omega$ in the lower spherical harmonics describing
the stellar structure and mode structure.  The uneven aspect of the lines
corresponding to the truncated calculations are due to interpolation errors.}
\label{fig:truncated}}
\end{tabular}
\end{figure}

The effects of the centrifugal force, on the other hand, increase with radial
order.  The reason for this, as explained in Paper~I, is that changes in the
stellar structure and the sound velocity profile causes modifications which are
roughly proportional to the frequencies.  For spherically symmetric changes in
a star's structure, we have $\Delta \ln \omega = - \Delta \ln \int_0^R dr/c$
(where $\Delta$ means the variations due to the change in stellar structure),
based on Tassoul's asymptotic formula \citep{Tassoul80}.  The same principle
applies for more complicated changes in the structure, such as those provoked
by the centrifugal force, but will have a more complicated mathematical
formulation.  One way of illustrating this is by plotting the ratios
$D_1/\omega_0$ (see \eq{perturbative}), which correspond to $\d \ln \omega/\d
\Omega^2$ calculated at $\Omega=0$ for $m=0$ modes.  We take the derivative
with respect to $\Omega^2$ since the effects of the centrifugal force begin at
$2^\mathrm{nd}$ order in $\Omega$.  In \fig{coeff}, we can see that these
ratios approach constant values as $n$ increases for a given $\l$. 
Furthermore, we have plotted these ratios both with and without the Coriolis
force, thereby demonstrating that this effect is entirely due to the
centrifugal force. For non-axisymmetric modes, the relevant ratios $(D_1 + m^2
D_2)/\omega_0$ also show the same behaviour.  This shows that the effects of
the centrifugal distortion is roughly proportional to the frequency.

It can then be expected that making errors on the effects of the centrifugal
force will also lead to differences proportional to the frequencies.  If we
look at panel~(b) of \fig{centrifugal} in which the Coriolis force has been
suppressed, we can see that the relative differences between perturbative and
complete calculations $\delta \omega/\omega$ actually increase with radial
order (at least for lower and moderate rotation rates).  It is an open question
whether or not these relative differences will approach an asymptotic limit as
the radial order increases like in \fig{coeff}.

The increase of perturbative errors with radial order may have important
implications for stars which pulsate in high radial overtones.  Examples of
these are solar type stars which typically pulsate with radial orders
between~$15$ and~$25$.  If we assume that perturbative errors scale with
frequency and frequency with radial order, we can estimate at what point
complete methods will be necessary to calculate the effects of rotation on such
modes.  The average limit for $n=25$ pulsation modes in $M = 1 \, M_\odot$,
$\Rp = 1 \, R_\odot$ stars would be $v_\mathrm{eq} = 25\,\kms$ ($45\,\kms$) for
COROT's primary (secondary, resp.) program.  This implies that non-perturbative
effects of rotation could be visible for moderate rotation rates.  Nonetheless,
direct comparisons between perturbative and complete calculations for high
order modes are necessary to confirm this conclusion.


In order to further understand the role of the centrifugal force in
perturbative errors, it is helpful to bear in mind the approximations which
result from applying perturbative methods.  First of all, the frequencies, the
stellar structure and the mode structure are all described by low degree
polynomials in $\Omega$.  This then leads to the following effects:  the
equilibrium structure is only described by $\l=0$ and $\l=2$ spherical
harmonics (for $2^\mathrm{nd}$ and $3^\mathrm{rd}$ order methods) and the
pulsation mode structure is also limited to a few spherical harmonics.  In
order to analyse the effects of using only a few spherical, we did highly
truncated numerical calculations, in which the equilibrium model has been
reduced to $2$ spherical harmonics and the pulsation modes to $4$ spherical
harmonics ($2$ poloidal + $2$ toroidal).  In \fig{truncated}, the black lines 
represent the relative differences between these truncated calculations and
complete ones.  These differences are significant, thus showing the need for
more spherical harmonics.  Nonetheless, we also plot in \fig{truncated} the
perturbative errors which are higher than the truncated calculations.  This
shows that including higher order terms in $\Omega$ in the contribution of even
the lowest degree spherical harmonics can improve results.


\subsection{Discussion}

As can be seen from previous sections, differences between perturbative
calculations and complete ones can be quite substantial.  This is problematic
because obtaining accurate results is crucial in asteroseismology.  Differences
between theoretical calculations and observed frequencies need to come from
differences between the stellar model and the star's actual structure rather
than from an approximate treatment of the effects of rotation.  Otherwise,
modifying the stellar structure so as to match the observed frequencies will
end up compensating errors in the calculation of frequencies instead of
improving the stellar model.  Moreover, large errors on frequency calculations
can lead to erroneous mode identifications, especially if proximity between
observed and theoretical frequencies is the only criteria used for establishing
such identifications.  Interestingly, establishing a correct mode
identification is one of the key difficulties in interpreting $\delta$ Scuti
stars \citep[\textit{e.g.}][]{Goupil05}.  Mode misidentification can occur
because frequencies are not in the same order in perturbative calculations as
in complete ones (in \fig{differences}, this can be seen by the dotted lines
for the $\l=1$, $2$ or $3$ modes which cross each other, thus indicating an
exchange of position between two frequencies) and because one can no longer
rely on the usual frequency patterns used in slowly rotating stars (see
Paper~I).  An erroneous mode identification then leads a false interpretation
of pulsation frequencies due to an incorrect understanding of the geometry of
the pulsation mode and the stellar regions which it probes.  This problem is
further aggravated by the fact that perturbative methods only give an
approximate idea of the structure of a given mode anyway.  Fully taking into
account the effects of rotation on stellar pulsation increases the likelihood
of obtaining a correct identification and gives a better understanding of mode
structure, especially when the rotation rate is high.  However, in order to
obtain such a mode identification, the underlying stellar model needs to be
sufficiently close to reality so as enable a successful matching between
theoretical predictions and observations.  It is possible that even then, mode
identification is uncertain due to multiple solutions which fit a set of
observed frequencies.

\section{Conclusion}

In this work we have explored some of the effects of rapid rotation on stellar
acoustic pulsations.  This was achieved thanks to a numerical method which
combines spheroidal geometry and spectral methods, as in Paper~I. Through a
detailed analysis, we have shown that our results have a $6$ to $7$ digit
accuracy.  This analysis included a discussion on the accuracy of the
underlying polytropic models, comparisons with \citet{S81,C84,CDM94} and
Paper~I, a test based on the variational principle and some tests on the
sensitivity of the results to the parameters of the numerical method.

In the future, the satellite COROT is expected to measure stellar oscillation
frequencies with a precision of $0.08\,\mathrm{\mu Hz}$ (primary targets) or
$0.6\,\mathrm{\mu Hz}$ (secondary targets). In the frequency range considered
in this paper, we find that for a $M=1.9\, M_\odot$, $R=2.3\,R_\odot$ star,
perturbative methods cease to be valid for COROT's primary (secondary) program
beyond $\vsini=50\,\kms$ ($\vsini=75\,\kms$, resp.).  At a rotation rate of
$0.59 \,\Omega_K$, the perturbative spectrum is very different from the one
based on complete calculations.  Therefore, any attempt to interpret stellar
pulsations using the perturbative approach at comparable rotation rates is
likely to fail. Using complete methods on the other hand increases the
likelihood of obtaining a correct mode identification, and gives an accurate
description of the structure of pulsation modes.  Both of these are crucial
when interpreting observed pulsation modes.

Further investigation has shown the dominant role of the centrifugal force in
modifying the frequency spectrum and causing perturbative errors.  This is
because while the effect of the Coriolis force decreases as the frequency
increases, the effect of the stellar deformation increases roughly
proportionally to the frequencies.  Therefore, the errors which arise from
perturbative descriptions of the centrifugal distortion are also amplified in
higher order modes.  As a result, it may be necessary to use complete methods
for moderately rotating stars which exhibit high order modes.

Some of the issues which were discussed in Paper~I and have yet to be discussed
for the present results include: an analysis of the regularities in the
oscillation spectrum at high rotation rates and a study of the visibility of
the different modes based on their structure.  These will be the subject of
forthcoming papers.  A few preliminary examinations have already confirmed some
of the conclusions given in Paper~I, such as a strong equatorial concentration
of mode structure at high rotation rates, or the transition from one frequency
spectrum organisation to another.

Future work includes working with more realistic models, and studying gravity
modes in spheroidal geometry.  The transition to more realistic models is
essential before being able to compare theoretical frequencies with
observations.  Coming up with realistic models that fully include the effects
of rotation and in particular the centrifugal distortion is no easy task, but
is the subject of active research \citep[\textit{e.g.}][]{Roxburgh04,
Jackson05, Rieutord05}.  Calculating the associated pulsation frequencies and
comparing them to observations will provide crucial information on stellar
structure and enable a better adjustment of these models.

The study of the effects of rapid rotation on g-modes is of interest for the
interpretation of $\gamma$ Doradus stars, which are g-mode pulsators and can be
rapid rotators.  Previous studies on the non-perturbative effects of the
Coriolis force on g-modes \citep{Dintrans99, Dintrans00} have revealed their
important role in altering the geometry and frequencies of these modes.  This
behaviour is entirely different from that of the high frequency acoustic modes
presented here.  It is then interesting to understand what the effects of the
centrifugal force will be on g-modes and how it will compare with the effects
of the Coriolis force.

\begin{acknowledgements}
The authors wish to thank Lorenzo Valdettaro for his contribution to the
numerical aspects of this work. Many of the numerical calculations were carried
out on the Altix 3700 of CALMIP (``CALcul en MIdi-Pyrénées'') and on the NEC
SX5 of IDRIS (``Institut du Développement et des Ressources en Informatique
Scientifique''), both of which are gratefully acknowledged.
\end{acknowledgements}

\bibliographystyle{aa}
\bibliography{biblio}
\addcontentsline{toc}{chapter}{References}
\appendix
\section{``Generalised'' Frobenius method}
\label{sect:Frobenius}

\subsection{Description}
The starting point in this method is the following equation:
\begin{equation}
\label{eq:Frobenius.main}
\frac{dY(x,y,z)}{dx} + \frac{1}{x} A(x,y,z)Y(x,y,z) = 0.
\end{equation}
This equation looks very much like a first order Frobenius equation except that
two other variables, $y$ and $z$, intervene in the different terms \citep[for a
description of the more traditional version of the Frobenius method, see][]
{BenderOrszag}. The quantity $Y(x,y,z)$ can be a scalar or a vector. The
operator $A(x,y,z)$ can include derivatives in the $y$ and $z$ directions and
needs to be analytic in the $x$ direction, so that we can write:
\begin{equation}
A(x,y,z) = \sum_{n=0}^{\infty} A_n(y,z) x^n.
\end{equation}

We then look for the behaviour of $Y(x,y,z)$ along the boundary $x=0$.  If
we develop $Y(x,y,z)$ in the following manner,
\begin{equation}
Y(x,y,z) = \sum_{n=0}^{\infty} Y_n(y,z) x^{n+\alpha},
\end{equation}
then we obtain the following zeroth order equation:
\begin{equation}
\label{eq:zeroth.order}
\alpha Y_0(y,z) + A_0(y,z) Y_0(y,z) = 0,
\end{equation}
where $\alpha$ is the leading power of $x$ in $Y(x,y,z)$.  Therefore, to obtain
$\alpha$, one needs to solve an eigenvalue problem in terms of the coordinates
$y$ and $z$, along the entire surface $x=0$. The remaining $Y_n$ are defined
through the following recurrence relation:
\begin{equation}
n \geq 1, \quad \left[(\alpha+n)Id + A_0 \right] Y_n = -
\sum_{k=0}^{n-1} A_{n-k} Y_k,
\label{eq:recurrence.Frobenius}
\end{equation}
where $Id(Y) \equiv Y$.  This series is defined only if for each $n \geq 1$,
the operator $\left[ (\alpha+n)Id + A_0 \right]$ is invertible. The next step
is then to search under what mathematical conditions the series defined by
\eq{recurrence.Frobenius} converges.  However, this step is quite complicated
and therefore beyond the scope of this paper.

\subsection{Application}
In our case, we are only interested in obtaining the leading behaviour of
our solutions near the surface.  Therefore we will only solve the zeroth order
equation (\eq{zeroth.order}) after having established the expressions
for $Y_0$ and $A_0$.  We start by defining $x=1-\zeta$ as the variable that
will be used in the Frobenius series.  The surface of the star then corresponds
to $x=0$ and its interior to positive values of $x$.

It is then necessary to choose a vector $Y$ so that
\eqtoeq{spheroidal.continuity}{spheroidal.Poisson} can be put in the form given
by \eq{Frobenius.main}.  This implies choosing the variables which
are differentiated once with respect to $x$.  Our choice is therefore
$Y=[\Pi,\uz,\G=\d_x \Psi,\Psi]^t$.  Having chosen the vector $Y$, it is then
necessary to find the associated system of equations, by eliminating the
variables $(b,\ut,\up)$, and then to extract the zeroth order equation (see
\eq{zeroth.order}).  In fact, it is much simpler to do both steps
simultaneously, given the complexity of  \eqtoeq{spheroidal.continuity}
{spheroidal.Poisson}.

Before giving the final result, it is important to point out that when $N$ is
not an integer, a mild singularity occurs on the surface of the star, due to
the presence of fractional powers in the enthalpy, starting with $x^{N+2}$
\citep{H01}.  This in fact invalidates the use of Frobenius series in its
present form from a strictly mathematical point of view, since these only use
integer powers of $x$. This problem can be solved by including fractional
powers in the solution, as is done in \citet{CDM94}, the lowest one being
$x^{N+1}$ (this is not in contradiction with \citet{CDM94}, for which the
variable $y_4$ contains $x^N$, because $y_4$ includes $\d_{xx}H$ in its
expression whereas our variables do not).  As a result, the zeroth order
equation remains unaffected and can therefore give the correct behaviour of the
solution near the surface:
\begin{equation}
\alpha
\left[ \begin{array}{c} \Pi_0 \\ \uz_0 \\ \G_0 \\ \Psi_0 \end{array} \right] + 
\left[
\begin{array}{cccc}
\displaystyle N \left( 1 - \frac{\gamma}{\Gamma_1} \right) &
\displaystyle \frac{N H_x}{\lambda \Lambda(1-\varepsilon)R_s^2}
\left(\frac{\gamma}{\Gamma_1}-1\right) &
0 & 0 \\
\displaystyle -\frac{\lambda\Lambda(1+N)(1-\varepsilon)R_s^2}{H_x \Gamma_1} &
\displaystyle \frac{1+N}{\Gamma_1} &
0 & 0 \\
0&0&0&0\\0&0&0&0\\
\end{array} \right]
\left[ \begin{array}{c} \Pi_0 \\ \uz_0 \\ \G_0 \\ \Psi_0 \end{array} \right]
=0.
\end{equation}

This equation is based on the following development of the enthalpy near the surface:
\begin{equation}
H(x,\theta) = H_x(\theta) x + \frac{1}{2}H_{xx}(\theta) x^2 + H_N(\theta) x^{N+2} +
o(x^2).
\end{equation}
The characteristic equation is $det(A_0-X.Id) =  X^4 - N X^3 = 0$. The
eigenvalues are therefore $\alpha = -N$ and $\alpha = 0$, the second value
being triply degenerate.  The first value is rejected because it leads to
solutions that diverge on the surface of the star.  The three remaining
eigensolutions are bounded near the surface, which is in complete agreement
with the results of \citet{HRW66}, who applied the Frobenius method to the
spherical case.  By choosing $\alpha = 0$, we also ensure that $\left[ (\alpha
+ n)Id + A_0 \right]$ is invertible for $n \geq 1$. These three bounded
solutions and any of their linear combinations satisfy the following analytical
constraint:
\begin{equation}
\label{eq:Frobenius.constraint}
\lambda \Pi_0 = \frac{H_x}{\Lambda (1 - \varepsilon) R_s^2} \uz_0,
\end{equation}
which is, in fact, equivalent to saying that $\delta p/\rho_0$ goes to zero on
the outer boundary (where $\delta p$ represents the Lagrangian variation of the
pressure).  We can use the previous results to establish the behaviour of
different quantities near the surface:
\begin{eqnarray}
p           & = & \O\left(x^{N}\right), \\
\rho        & = & \O\left(x^{N-1}\right), \\
\uz         & = & \O(1), \\
\ut         & = & \O(1), \\
\up         & = & \O(1), \\
\Psi        & = & \O(1).
\end{eqnarray}
\Eq{Frobenius.constraint} shows that $\delta p/H^N =  o\left(1\right)$.
The next relevant power in a power series expansion of $\delta p/H^N$ is
$x^1$ (this remains true even when $N$ is not an integer since the first
fractional power of $\delta p/H^N$ is $N+1$). By applying the equation
$\delta p = c_o^2 \delta \rho$, it can also be shown that  $\delta
\rho/H^{N-1} = \O(x)$. As a result we obtain the following behaviour for
both Lagrangian perturbations:
\begin{eqnarray}
\delta p    & = & \O\left(x^{N+1}\right), \\
\delta \rho & = & \O\left(x^{N}\right).
\end{eqnarray}

The results on $\rho$, $p$, $\delta \rho$, $\delta p$ are interesting when we
consider the equilibrium model.  Since $\rho_o \propto H^N$ and $P_o \propto
H^{N+1}$, we deduce that the leading behaviour of the equilibrium density and
pressure are $\rho_o = \O \left( x^{N} \right)$ and $P_o = \O \left( x^{N + 1}
\right)$, respectively.  This implies that the ratio of the Eulerian density
perturbation to the equilibrium density $(\rho/\rho_o)$ and the corresponding
ratio for pressure both become unbounded as one approaches the surface of the
star.  This is problematic because the sum $\rho_o + A \, \rho \, \cos (\omega
t)$ (which corresponds to the total density) will periodically reach negative
values close to the surface of the star for any non-zero amplitude $A$. 
However, the ratio of the Lagrangian density perturbation to the equilibrium
density remains bounded as one approaches the surface, and the same applies to
the pressure.  This suggests that a Lagrangian description is physically more
appropriate.

\section{Projection onto the spherical harmonic base}
\label{sect:harmonic}

\subsection{Integral operators}
In order to project the fluid equations onto the harmonic basis, it is
necessary to define a number of integral operators.  The prototype to one of
these operators is as follows:
\begin{equation}
\Jllm{G}(\zeta) = \int \!\!\!\! \int_{4 \pi} G(\zeta,\theta)
           \Ntlmp(\theta,\phi) \left\{ \Ylm(\theta,\phi) 
           \right\}^* \mathrm{d}\Omega,
\end{equation}
where $d\Omega = \sin\theta \mathrm{d} \theta \mathrm{d} \phi$, $G$ is an
arbitrary function, $x^*$ is the complex conjugate of $x$ and $\Jllm{.}$ is the
operator.  $\Jllm{G}$ is a two-dimension array of indexes $\l$ and $\l'$ (the
value of $m$ is fixed) composed of functions depending on $\zeta$ only.  The
remaining operators are given in the following table:
$$
\begin{array}{c|c|c|c}
 & \left\{ \Ylm \right\}^* & \left\{ \Ntlm \right\}^* & \left\{ \Nplm \right\}^* \\
\hline
\Ylmp       & \Illm{G} & \Jllmc{G} & \Kllmc{G} \\
\Ntlmp      & \Jllm{G} & \Lllm{G}  & \Mllmc{G} \\
\Nplmp      & \Kllm{G} & \Mllm{G}  & \Nllm{G} 
\end{array}
$$

If $G$ is a real function than $\Illm{G}$, $\Jllm{G}$, $\Jllmc{G}$, $\Lllm{G}$
and $\Nllm{G}$ are all real functions whereas $\Kllm{G}$, $\Kllmc{G}$,
$\Mllm{G}$, and $\Mllmc{G}$ are purely imaginary. There are symmetries between
some of these operators: for example $\Jllm{G^*} = \left\{
J\!c_{\l'\l}^m \left( G \right) \right\}^*$.  The same applies for $\Kllm{G}$
and $\Kllmc{G}$, and for $\Mllm{G}$ and $\Mllmc{G}$.

In order to calculate these integrals, we use Gauss' quadrature. This gives
accurate integrals when  $G$ is a ``polynomial'' of $\cost$ (the coefficients
of the polynomial can depend on $\zeta$), for the operators $\Illm{G}$,
$\Lllm{G}$, $\Mllm{G}$, $\Mllmc{G}$ and $\Nllm{G}$. For the operators
$\Jllm{G}$, $\Kllm{G}$, $\Jllmc{G}$ and $\Kllmc{G}$, $G$ needs to be of the
form $\sint P(\cost)$ where $P$ is a polynomial.  These integrals are
calculated with $\Lres$ collocation points, where $\Lres$ is generally
greater than $\Lmax$, the harmonic resolution of the pulsations.

Having defined the different integral operators, it is now possible to give
explicitly the fluid equations projected onto the spherical harmonic basis.  In
what follows, we have used the following conventions:
\begin{equation}
\Illm{G} \ulmp  \equiv \sum_{\l'} \Illm{G} \ulmp, \qquad
\LNllm{-}{+}{G} \equiv -\Lllm{G}+\Nllm{G}.
\end{equation}
It is also worth pointing out that in the following matrices, the summation
on $\l'$ applies to an entire line of the matrix. For example,
\begin{equation}
\left[ \begin{array}{lr}
+\Illm{A}-\Jllm{B} & \ulmp \\
-\Kllm{C}+\Nllm{D} & \vlmp
\end{array} \right],
\end{equation}
is equivalent to:
\begin{equation}
\sum_{\l'=|m|}^{L}  \left\{\Illm{A}-\Jllm{B}\right\}   \ulmp 
+\sum_{\l'=|m|}^{L} \left\{-\Kllm{C}+\Nllm{D}\right\}  \vlmp.
\end{equation}

\subsection{Continuity equation}

\begin{equation}
\lambda	 \blm =
\left[
  \begin{array}{lr}
  -\Illm{\frac{\zeta^2 H}{r^2\rz}}         & \dz \ulmp \\
  -\Illm{\frac{2\zeta H}{r^2\rz}
  +\frac{\zeta^2 N \Hz}{r^2\rz}}           & \ulmp \\
  +\Illm{\frac{\l'(\l'+1)\zeta H}{r^2\rz}}
  -\Jllm{\frac{\zeta N \Ht}{r^2\rz}}       & \vlmp \\
  -\Kllm{\frac{\zeta N \Ht}{r^2\rz}}       & \wlmp
  \end{array}
\right],
\end{equation}
where we have made use of the following identities:
\begin{eqnarray}
\label{eq:identity}
-\l(\l+1)\Ylm &=& \dtt \Ylm + \cot \theta \dt \Ylm + \frac{1}{\sin^2 \theta}
                  \dpp \Ylm, \\
0             &=& \dt \Nplm + \cot \theta \Nplm - \frac{1}{\sin \theta}
                     \dphi \Ntlm .
\end{eqnarray}

\subsection{Adiabatic energy equation}

\begin{equation}
\lambda \left( \Pilm - \frac{\Gamma_1}{(N+1)\Lambda} \blm \right) = 
\left( \frac{\Gamma_1}{\gamma} -  1 \right) \left[
\begin{array}{lr}
+\Illm{\frac{\zeta^2 \Hz}{\Lambda r^2 \rz}} & \ulmp \\
+\Jllm{\frac{\zeta   \Ht}{\Lambda r^2 \rz}} & \vlmp \\
+\Kllm{\frac{\zeta   \Ht}{\Lambda r^2 \rz}} & \wlmp
\end{array} \right].
\end{equation}

\subsection{Poisson's equation}
\begin{equation}
0 = \left[
\begin{array}{lr}
+\Illm{\frac{r^2+\rt^2}{\rz^2}}           & \dzz \Psilmp \\
+\Illm{r^2 \cz} -2 \Jllm{\frac{\rt}{\rz}} & \dz \Psilmp \\
- \l(\l+1)                                & \Psilm \\
- \Illm{r^2H^{N-1}}                     & \blmp 
\end{array} \right],
\end{equation}
where we have made use of the \eq{identity}.

\subsection{Euler's equations}
\begin{equation}
\lambda \left[
\begin{array}{lr}
+ \Illm{\frac{\zeta^2 \rz H}{r^2}} & \ulmp \\
+ \Jllm{\frac{\zeta\rt H}{r^2}}    & \vlmp \\
+ \Kllm{\frac{\zeta\rt H}{r^2}}    & \wlmp
\end{array} \right] = 
\left[ \begin{array}{lr}
+ \Kllm{\frac{2 \Omega H \zeta \sint}{r}} & \vlmp \\
- \Jllm{\frac{2 \Omega H \zeta \sint}{r}} & \wlmp \\
- \Illm{H}                                & \dz \Pilmp \\
- \Illm{H}                                & \dz \Psilmp \\
- \Illm{N \Hz}                          & \Pilmp \\
+ \Illm{\frac{\Hz}{\Lambda}}              & \blmp
\end{array} \right],
\end{equation}

\begin{equation}
\lambda \left[
\begin{array}{lr}
+ \Jllmc{\frac{\zeta\rt}{r^2}}                            & \ulmp \\
+ \Lllm{\frac{r^2+\rt^2}{r^2\rz}} + \Nllm{\frac{1}{\rz}}  & \vlmp \\
+ \Mllm{\frac{r^2+\rt^2}{r^2\rz}} - \Mllmc{\frac{1}{\rz}} & \wlmp
\end{array} \right] = 
\left[ \begin{array}{lr}
- \Kllmc{\frac{2\Omega \zeta \sint}{r}}                   & \ulmp \\
  \MMcllm{+}{-}{\frac{2\Omega(\rt \sint + r \cost)}{r\rz}}& \vlmp \\
  \LNllm{-}{-}{\frac{2\Omega(\rt \sint + r \cost)}{r\rz}} & \wlmp \\
  \LNllm{-}{-}{\frac{1}{\zeta}}
- \Jllmc{\frac{N\Ht}{\zeta H}}                          & \Pilmp \\
  \LNllm{-}{-}{\frac{1}{\zeta}}                           & \Psilmp \\
+ \Jllmc{\frac{\Ht}{\Lambda \zeta H}}                     & \blmp
\end{array} \right],
\end{equation}

\begin{equation}
\lambda \left[
\begin{array}{lr}
+ \Kllmc{\frac{\zeta\rt}{r^2}}                            & \ulmp \\
+ \Mllmc{\frac{r^2+\rt^2}{r^2\rz}} - \Mllm{\frac{1}{\rz}} & \vlmp \\
+ \Nllm{\frac{r^2+\rt^2}{r^2\rz}} + \Lllm{\frac{1}{\rz}}  & \wlmp
\end{array} \right] =
\left[ \begin{array}{lr}
+ \Jllmc{\frac{2\Omega \zeta \sint}{r}}                   & \ulmp \\
  \LNllm{+}{+}{\frac{2\Omega(\rt \sint + r \cost)}{r\rz}} & \vlmp \\
  \MMcllm{+}{-}{\frac{2\Omega(\rt \sint + r \cost)}{r\rz}}& \wlmp \\
  \MMcllm{+}{-}{\frac{1}{\zeta}}
- \Kllmc{\frac{N\Ht}{\zeta H}}                          & \Pilmp \\
  \MMcllm{+}{-}{\frac{1}{\zeta}}                          & \Psilmp \\
+ \Kllmc{\frac{\Ht}{\Lambda \zeta H}}                     & \blmp
\end{array} \right].
\end{equation}

\section{The variational test}
\label{sect:variational}

The present formulation of the variational test is the same as that of
\citet{Unno89}, apart from the following differences: we use the velocity
rather than the displacement, hence the extra time derivatives; the star's
volume is no longer spherical; the integral on the gravity wave energy is
based on the effective gravity and uses the local vertical direction rather
than $\er$; the integral on the gravitational potential energy has been
extended to infinite space.

The different resultant integrals are given by the following explicit formulas
and are calculated numerically using Gauss' quadrature in the angular direction
and a spectral expansion in the radial direction (we
use a radial resolution of $101$ points and an angular resolution of $200$
points):
\begin{eqnarray}
\int_V \rho_o \| \vect{v} \|^2 dV &=& \int_V H^{N} \left[
   |\uz|^2 \frac{\zeta^4}{r^4} + |\ut|^2 \frac{\zeta^2(r^2+\rt^2)}{r^4 \rz^2}
  +|\up|^2\frac{\zeta^2}{r^2\rz^2} + 2 \Re \left\{ \left(\uz\right)^*\ut\right\}
   \frac{\zeta^3\rt}{r^4\rz} \right] dV, \\
\int_V \rho_o N_o^2 |\vect{v} \cdot \vect{e}_g |^2 dV &=& \int_V \frac{N H^{N-1}}
  {\Lambda} \left( 1 - \frac{\gamma}{\Gamma_1} \right) \frac{\zeta^2}{r^4 \rz^2}
  \left| \zeta \uz \dz H + \ut \dt H \right|^2 dV \\
\int_V \frac{|p|^2}{\rho_o c_o^2} dV &=& \int_V \frac{(N+1)\Lambda H^{N-1}}
          {\Gamma_1}|\Pi|^2 dV, \\
\int_V \rho_o \vect{\Omega} \cdot \left( \vect{v}^* \times \vect{v}\right) dV &=& 
       2i\Omega \int_{V} H^N \left[ \left( \frac{\cost}{\rz} + \frac{\rt\sint}{r\rz}
       \right) \frac{\zeta^2\left(u_r^{\theta} u_i^{\phi} - u_r^{\phi}
       u_i^{\theta} \right)}{r^2\rz} + \frac{\zeta^3 \sint \left(
       u_r^{\zeta}u_i^{\phi}-u_r^{\phi}u_i^{\zeta} \right)}{r^3\rz} \right] dV, \\
\int_{V\,\,\mathrm{or}\,\,V_2} \| \grad \Psi \|^2 dV &=& 
        \int_{V\,\,\mathrm{or}\,\,V_2} \frac{r^2 + \rt^2}{r^2\rz^2} \left|
        \dz \Psi \right|^2  + \frac{1}{r^2} \left| \dt \Psi \right|^2
        +\frac{1}{r^2 \sint^2} \left| \dphi \Psi \right|^2 - \frac{2\rt}{r^2\rz}
         \Re \left( \dz \Psi^* \dt \Psi \right) dV,
\end{eqnarray}
where $dV=r^2 |\rz| \sint d\theta d\zeta d\phi$, $u_r^{\zeta} = \Re\left(\uz\right)$,
$u_i^{\zeta} = \Im\left(\uz\right)$ etc.  For the integral on the gravitational
potential, it is useful to decompose infinite space into three domains: $V \cup
V_2 \cup V_3 = V_{\infty}$.  $V$ is the volume of the star, $V_2$ is the volume
comprised between the star and the sphere of radius $2$, and $V_3$ is the space
outside the sphere of radius $2$ (see \fig{domains}).  The integral on the
first two domains is given by the expression above. For the third domain, it is
based on the spherical harmonic decomposition of the gravitational potential. 
In empty space, a gravitational potential will take on the following form as it
obeys the equation $\lapl \Psi = 0$ and vanishes towards infinity:
\begin{equation}
\label{eq:phi.decomposition}
\Psi  = \sum_{\l} \Psilm \Ylm = \sum_{\l} \frac{A^{\l}}{r^{\l+1}} \Ylm,
\end{equation}
where the $A^{\l}$ are constants.  This form of $\Psi$ then leads to the
following expression:
\begin{equation}
\int_{V_3} \| \grad \Psi \|^2 dV = \sum_{\l} \frac{|A^{\l}|^2(\l+1)}
{r_{ext}^{2\l+1}} = \sum_{\l} r_{ext} (\l + 1) \left|\Psilm(r_{ext})\right|^2,
\end{equation}
where $r_{ext} = 2$ is the radius of the inner sphere of $V_3$.  This expression
corresponds to the surface integral of \citet{Unno89}.
\end{document}